\documentclass[aip,apm,reprint,amsmath,amssymb,twocolumn]{revtex4-1}

\usepackage{setspace}

\usepackage{graphicx}% Include figure files
\usepackage{times}

\begin{document}

%\preprint{}

\title{Dynamics in hybrid complex systems of switches and oscillators}

\author{Dane Taylor}
\email{dane.taylor@colorado.edu} 
\affiliation{Department of Applied Mathematics, University of Colorado, Boulder, CO 80309,~USA}
%\affiliation{Department of Mathematics, University of North Carolina, Chapel Hill, NC 27599,~USA}
%\affiliation{Statistical and Applied Mathematics Sciences Institute, Research Triangle Park, NC 27709,~USA}

\author{Elana J. Fertig}
\email{ejfertig@jhmi.edu}
\affiliation{Oncology Biostatistics, John Hopkins University, Baltimore, MD 21205,~USA}

\author{Juan G. Restrepo}
\email{juanga@colorado.edu}
\affiliation{Department of Applied Mathematics, University of Colorado, Boulder, CO 80309,~USA}

\date{\today}

%%%%%%%%%%%%%%%%%%%%%%%%%%%%%%%%%%%%%%%%%%%%%%%%%%%%%%%%
\begin{abstract}

While considerable progress has been made in the analysis of large systems containing a single type of coupled dynamical component (e.g., coupled oscillators or coupled switches), systems containing diverse components (e.g., both oscillators and switches) have received much less attention. We analyze large, hybrid systems of interconnected Kuramoto oscillators and Hopfield switches with positive feedback. In this system, oscillator synchronization promotes switches to turn on. In turn, when switches turn on they enhance the synchrony of the oscillators to which they are coupled. Depending on the choice of parameters, we find theoretically coexisting stable solutions with either (i) incoherent oscillators and all switches permanently off, (ii) synchronized oscillators and all switches permanently on, or (iii) synchronized oscillators and switches that periodically alternate between the on and off states. Numerical experiments confirm these predictions. We discuss how transitions between these steady state solutions can be onset deterministically through dynamic bifurcations or spontaneously due to finite-size fluctuations. 

\end{abstract}
%%%%%%%%%%%%%%%%%%%%%%%%%%%%%%%%%%%%%%%%%%%%%%%%%%%%%%%%
\maketitle

{\bf
%Theory for complex systems \cite{complex_syst} and complex networks\cite{complex_net} are of broad interest across scientific disciplines such as physics  \cite{Millenium,Josephson,Lasers, Switch_electronics}, social behavior \cite{FireFly,Animal,Clapping}, and physiology \cite{Pacemaker,Circadian,Kuramoto1,Hopfield,bool_gene,Chem_switch}. 
Although extensive theoretical progress has been made in understanding collective behavior in large systems containing a single type of component (such as a switch \cite{bool_gene} or oscillator \cite{sync}), there has been less development for diverse systems containing more than one type of component. However, many complex systems are composed of various types of units \cite{multiplex,interactome,interconnected,motif,Tyson, yeast, elana}. 
%We analyze the macroscopic dynamics arising in large hybrid systems containing both switches and oscillators. Our work is motivated by the observation that diversity is ubiquitous in complex systems \cite{multiplex,interactome,interconnected,motif, yeast, elana}. 
For example, the system-wide dynamics of the yeast cell cycle may be modeled as a system of coupled switches and oscillators \cite{yeast, elana}.
 Extending the numerical work of Ref.\cite{elana}, we study interconnected Hopfield switches \cite{Hopfield} and Kuramoto oscillators \cite{Kuramoto1} with positive feedback. We find three steady state solutions that may coexist: (i) the Incoherent-Off (I-Off) state in which the oscillators are incoherent and all switches are permanently off, (ii) the Synchronized-On (S-On) state in which the oscillators synchronize and all switches remain on, and (iii) the Synchronized-Periodic (S-P) state in which the oscillators synchronize and the switches periodically turn on and off. Numerical experiments confirm our predictions for these steady state solutions and the transitions between them. 
 %Moreover, these transitions are classified as deterministic (if onset by a dynamic bifurcation caused by the temporal variation of a parameter \cite{DynBif}) or spontaneous (if stochastically onset by finite-size fluctuations \cite{daido,vary_freq,macro}). 
Our model demonstrates how the interplay between different units can result in rich dynamics. 
}

%%%%%%%%%%%%%%%%%%%%%%%%%%%%%%%%%%%%%%%%%%%%%%%%%%%%%%%%
\section{Introduction}\label{intro}
%%%%%%%%%%%%%%%%%%%%%%%%%%%%%%%%%%%%%%%%%%%%%%%%%%%%%%%%

The interdisciplinary nature of modern scientific research has demonstrated the pervasive need of theory for complex systems \cite{bool_gene,sync} and complex networks \cite{complex_net}. Of particular interest are large systems involving interconnected components, such as interacting neurons, genes, or people, that are responsible for outcomes in the larger system that they compose.
%Because the emergent behavior arising for such complex systems is not easily predicted from the behavior of individual components, the development of mathematical frameworks is essential for understanding and predicting system-wide behavior.
 % 
 Significant advances have been made for complex systems containing a single type of component. For example, models of synchronization of oscillators have been used to study collective phenomena in physics (e.g., pedestrian bridges \cite{Millenium}, Josephson junction circuits \cite{Josephson}, and lasers \cite{Lasers}), social behavior (e.g., flashing of fireflies \cite{FireFly}, animal flocking \cite{Animal}, and audiences clapping \cite{Clapping}) and physiology (e.g., circadian rhythms \cite{Circadian} and chemical oscillators \cite{Kuramoto1}). Similarly, interacting switches have been used to investigate gene expression \cite{bool_gene}, neural processing \cite{Hopfield}, electronic circuits \cite{Switch_electronics}, and chemical reactions \cite{Chem_switch}. In spite of these advances, the investigation of systems with diversity remains an open topic at the forefront of complex systems research \cite{multiplex,interactome,interconnected,motif, Tyson, yeast,elana}. 
 
 %For example, several forms of diversity include:
%
%(a) {\it multiplex networks} \cite{multiplex} containing diverse types of interactions (e.g., interactome networks describing relations between genes, metabolites, and proteins \cite{interactome}); 
%
%(b) {\it interconnected networks} \cite{interconnected} in which networks with differing dynamics interact (e.g., the physical power grid controlled by a communication system and transmission switching \cite{power}); and 
%
%© {\it motif networks} \cite{motif} whose macroscopic dynamics relate to the accumulation of diverse dynamics that naturally arise from the structure of overrepresented subgraphs (e.g., gene interaction networks whose motifs often yield switch-like or oscillatory dynamics \cite{motif2}).

Recently, a model was developed to study hybrid systems composed of coupled switches and oscillators \cite{elana}. 
The hybrid model recapitulated the system-wide dynamics of the yeast cell cycle, while demonstrating that small perturbations in the network topology result in cancer-like limitless activation of the cell cycle machinery. 
Motivated by these results, and by the observation that such hybrid systems allow the investigation of various forms of diversity \cite{multiplex,interactome,interconnected,motif, Tyson},
%[e.g., types (a-c)]
 we extend these numerical results and analyze theoretically the dynamics of a hybrid system of coupled oscillators and switches. 
%simulations to further develop a formal theoretical framework to understand hybrid systems. 
Our analysis utilizes the paradigmatic frameworks of Kuramoto oscillators \cite{Kuramoto1} and Hopfield switches \cite{Hopfield} to investigate stable solutions arising for large systems with positive feedback, i.e., oscillator synchronization promotes switches to turn on and when switches turn on they enhance the synchrony of the oscillators to which they are coupled. 
As a result, we find coexisting, parameter-dependent, stable solutions with (i) incoherent oscillators and all switches permanently off, (ii) synchronized oscillators and all switches permanently on, or (iii) synchronized oscillators and switches that periodically alternate between the on and off states. 
%Numerical experiments confirm these states, while suggesting that transitions between these steady state solutions can be onset deterministically through dynamic bifurcations or spontaneously due to finite-size fluctuations. 
Numerical experiments show that, in addition to deterministic transitions between states due to slow parameter changes, there are stochastic transitions between states mediated by finite-size fluctuations.

The remainder of this paper is organized as follows: In Sec.~\ref{sect:model} we introduce our model, discuss the parameter ranges of interest, and provide an overview of the dynamics to be studied. In Sec.~\ref{sect:analysis} we analyze three types of steady state solutions: an Incoherent-Off state (Sec.~\ref{sect:analysis_IO}), a Synchronized-On state (Sec.~\ref{sect:analysis_SO}), and a Synchronized-Periodic state (Sec.~\ref{sect:analysis_SP}). These results are validated by numerical experimentation in Sec.~\ref{sect:num}, where we explore transitions between steady state solutions (Sec.~\ref{sect:num_trans}) and investigate the relaxation of assumptions made in our analysis (Sec.~\ref{sect:num_relax}). Conclusions are drawn in Sect.~\ref{sect:conclusion}.

%%%%%%%%%%%%%%%%%%%%%%%%%%%%%%%%%%%%%%%%%%%%%%%%%%%%%%%%
\section{Model}\label{sect:model}
%%%%%%%%%%%%%%%%%%%%%%%%%%%%%%%%%%%%%%%%%%%%%%%%%%%%%%%%

\begin{figure}[t]
\centering
\includegraphics[width=6cm]{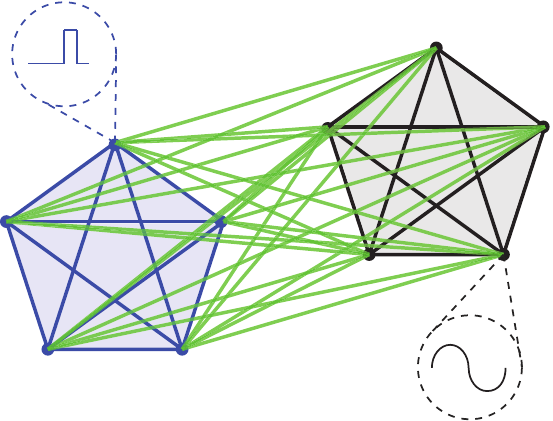} 
\caption{(Color online) An all-to-all network of phase oscillators (black nodes and links on right) is connected with an all-to-all network of switches (blue nodes and links on left) by connecting each node of a given network with all nodes in the other network (green links).
}
\label{fig:all-to-all}
\end{figure}

As an initial step toward analyzing hybrid models, we consider networks with all-to-all interactions, where each oscillator (or switch) is coupled to all other oscillators and switches, as illustrated in Fig.~\ref{fig:all-to-all}. The effect of network topology on hybrid systems will be explored in future research. To further facilitate our exploration, we focus our attention on interactions between Kuramoto phase oscillators \cite{Kuramoto1} and Hopfield switches \cite{Hopfield}, which respectively represent paradigmatic models for coupled oscillators and switches. 

Beginning with the Kuramoto phase oscillators \cite{Kuramoto1}, each oscillator $n=1,2,\dots,N$ is identically coupled to all the others by
\begin{eqnarray}
 \dot\theta_n  &=& \omega_n + \frac{k^{}}{N} \sum^N_{l=1} \sin(\theta_l-\theta_n), %\\
\label{eq:dtheta_0}
\end{eqnarray}
where $\theta_n(t)$ represents the phase of oscillator $n$ at time $t$, $\omega_n$ is oscillator $n$'s intrinsic frequency randomly chosen from a distribution $\Omega(\omega)$, and $k^{}(t)$ is the strength of coupling, which adapts to allow the switches to influence the oscillators. 
Recently, there has been much interest in adaptive dynamics of parameters in Eq.~\eqref{eq:dtheta_0}, including models that allow adaptation of the oscillator frequencies \cite{vary_freq}, coupling strength \cite{vary_K,macro}, or network structure \cite{vary_net}.

\begin{figure*}[t]
\centering
\includegraphics[width=18cm]{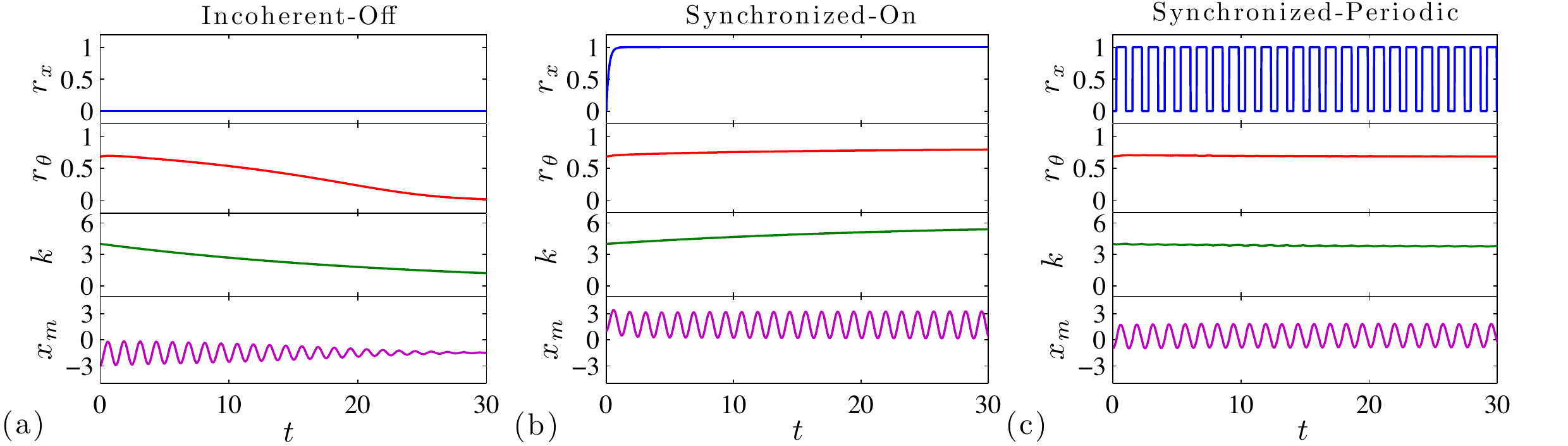} 
\caption{ (Color online) 
Time series are shown for simulations of Eqs.~(\ref{eq:system1}-\ref{eq:system3}) for $K^{\theta}=10$, $K=6$, $\tau=25$, $K^{x}=3.2$, $\eta=1.5$, $\Delta=1$, $\omega_0=5$, and $\beta_m=\beta=0$ with initial conditions $r_\theta(0)=0.7$, $k^{}(0) =4$, and three different distributions for $\{x_m(0)\}$. 
(a) For $\{x_m(0)\}$ values chosen with mean -1 and standard deviation 1, the system is initialized in the basin of attraction for the I-Off state. 
(b) For $\{x_m(0)\}$ values with mean 3 and standard deviation 1, the system is initialized in the basin of attraction for the S-On state. 
(c) For $\{x_m(0)\}$ values with mean 1 and standard deviation 1, the system is initialized in the basin of attraction for the S-P state. 
}
\label{fig:Identical_states}
\end{figure*}

We next consider a system of $M$ uniformly coupled Hopfield switches \cite{Hopfield}. In this model, without coupling to oscillators, each switch $m=1,2,\dots,M$ has in internal parameter $x_m$ that evolves as
\begin{eqnarray}
\dot x_m  &=& -x_m-\-\eta + \frac{K^{x}}{M} \sum_{l=1}^M \tilde x_l \label{eq:dx_0},%\\
\end{eqnarray}
where $K^{x}$ represents the strength of interaction between switches and $\tilde x_m$ corresponds to an external variable through which switch $m$ can interact with other switches. While the internal variables $\{x_m\}$ are allowed to evolve continuously, the external variables $\{\tilde x_m\}$ are defined piecewise based on the internal variables and may be taken to represent a highly sensitive variable. For each switch $m$, we have that $\tilde x_m=1$ ($\tilde x_m=0$) for $x_m>0$ ($x_m\le 0 $) and the switch is said to be in the ``on'' (``off'') state. Finally, the parameter $\eta$ can be interpreted as a threshold: if the last term in Eq.~\eqref{eq:dx_0} is larger than $\eta$ for a long enough time, switch $m$ will turn on.

We now introduce our mechanism for interconnectivity between oscillators and switches. As previously mentioned, the switches influence the oscillators through an adaptive oscillator coupling strength $k^{}$, which evolves according to the following relaxation model,
\begin{eqnarray}
\tau \dot k^{} &=& - k^{} + \frac{K}{M} \sum_{l=1}^M \tilde x_l  . \label{eq:dk_0}
\end{eqnarray}
Here $K$ determines the maximal coupling strength and $\tau$ controls the timescale for adaptation. To couple the switches to the oscillators, we consider an additional coupling term in
%propose the following replacement for Eq.~\eqref{eq:dx_0},
\begin{eqnarray}
\dot x_m  &=& -x_m-\eta + \frac{K^{x}}{M} \sum_{l=1}^M \tilde x_l + \frac{K^{\theta} }{N}\sum_{l=1}^N  \sin(\theta_l- \beta_m). \label{eq:dx} \nonumber\\
\end{eqnarray}
Note that in addition to interacting with other switches as described by Eq.~\eqref{eq:dx_0}, each switch is also influenced by each oscillator's phase. Specifically, the effect of the last term in Eq~\eqref{eq:dx} is that oscillator $l$ will promote the turning on of switch $m$ when its phase $\theta_l$ is close to $\beta_m+\pi/2$. Phase lags $\{\beta_m\}$ are randomly chosen from a distribution $B(\beta)$. 
%The introduction of phase lag heterogeneity is motivated by the observation that many natural systems contain variation, whether due to interaction delays \cite{delay} or spacial separation \cite{space}. 
In this paper several distributions $B(\beta)$ will be considered.
%, and in Sec. \ref{sect:num} we study the important role that heterogeneity can play in stabilizing the macroscopic dynamics of this system.

Having defined our hybrid model we now simplify the notation by adopting order parameters to measure collective behavior. The extent of synchrony may be measured with an order parameter $r_\theta$ and mean-field phase $\psi$, which are defined by $r_\theta e^{i\psi}=N^{-1}\sum_{n=1}^Ne^{i\theta_n}$. Similarly, we denote by $r_x=M^{-1}\sum_{m=1}^M\tilde x_m$ the fraction of switches in the on state. It follows that our model is given by the following system of $M+N+1$ equations
\begin{eqnarray}
\dot x_m  &=& -x_m-\eta + K^{x} r_x + K^{\theta} r_\theta\sin(\psi- \beta_m), \label{eq:system1}\\
 \dot\theta_n  &=& \omega_n + k^{}r_\theta \sin(\psi-\theta_n),\label{eq:system2}\\
\tau \dot k^{} &=& - k^{} + Kr_x  .\label{eq:system3}
\end{eqnarray}

Before concluding, we point out that although our model, Eqs.~(\ref{eq:system1}-\ref{eq:system3}), is similar to the hybrid model numerically studied by M. R. Francis and E. J. Fertig \cite{elana}, there are several important distinctive features:
First, while both models propose adding a new term to Eq.~\eqref{eq:dx_0}, the addition in the hybrid model of Ref.\cite{elana} was instead piecewise-defined to be 1 for $\theta_l\in[0,\pi]$ and 0 for $\theta_l\not\in[0,\pi]$. The new term, $\sin(\theta_l-\beta_m)$, has the same qualitative effect while being analytically tractable and preserving the continuity of the original Kuramoto model \cite{Kuramoto1}. 
Second, whereas Eqs.~(\ref{eq:system1}-\ref{eq:system3}) allow switches to affect oscillators through an adaptive coupling constant $k^{}$, the hybrid model of Ref.~\cite{elana} implements this interconnectivity instead by allowing the oscillators' intrinsic frequencies to adapt. We highlight this difference by offering the following interpretation for the effect of switches turning off on the oscillators: Whereas switches turning off under Eqs.~(\ref{eq:system1}-\ref{eq:system3}) may be interpreted as removing the coupling between oscillators, the turning off of switches in the hybrid model of Ref.~\cite{elana} causes oscillators' phases to freeze, in effect removing their ``oscillatory'' property.
Therefore, although an important advantage of the present model is analytical tractability, it is expected that both models will be relevant for various applications. The appropriate model should be selected, for example, based upon the physical structure of the network components \cite{motif,Tyson}.
Despite these differences, we find many similarities between the models' dynamics and thus the previous numerical experiments \cite{elana} will help guide our analysis.

%%%%%%%%%%%%%%%%%%%%%%%%%%%%%%%%%%%%%%%%%%%%%%%%%%%%%%%%
\subsection{Parameter choices}\label{sect:param}
%%%%%%%%%%%%%%%%%%%%%%%%%%%%%%%%%%%%%%%%%%%%%%%%%%%%%%%%

The free parameters in Eqs.~(\ref{eq:system1}-\ref{eq:system3}) are the distributions $\Omega(\omega)$ and $B(\beta)$ as well as the variables $K$, $K^{x}$, $K^{\theta}$, $\tau$, and $\eta$. 
We will focus on the case in which all oscillator-oscillator interactions are attractive, requiring $\tau,K>0$.  Moreover, oscillators following Eq.~\eqref{eq:dtheta_0} are well known to begin to synchronize when the coupling strength $k^{}>0$ is larger than some critical value $K_0>0$, which depends on the distribution of frequencies $\Omega(\omega)$ \cite{}. Therefore, to allow for the possibility of synchrony, we only consider values $K>K_0$. We will also only consider positive switch-switch and switch-oscillator interactions, which respectively requires $K^{x},K^{\theta}>0$. Finally, to preserve the bistability property of individual switches we require $\eta>0$. More specific choices will be discussed in Sec.~\ref{sect:num}

%%%%%%%%%%%%%%%%%%%%%%%%%%%%%%%%%%%%%%%%%%%%%%%%%%%%%%%%
\subsection{Overview of dynamics}\label{sect:overview}
%%%%%%%%%%%%%%%%%%%%%%%%%%%%%%%%%%%%%%%%%%%%%%%%%%%%%%%%

We will focus on three macroscopic states for our system: 
% that are characterized by non-trivial interactions between switches and oscillators: 
(i) the Incoherent-Off (I-Off) state in which the oscillators are incoherent and the switches all remain in the off state (Sec.~\ref{sect:analysis_IO}); 
(ii) the Synchronized-On (S-On) state in which the oscillators synchronize and the switches all remain in the on state (Sec.~\ref{sect:analysis_SO}); 
and (iii) the Synchronized-Periodic (S-P) state in which the oscillators synchronize and each switch periodically fluctuates between the on and off states (Sec.~\ref{sect:analysis_SP}). Example dynamics of system variables approaching these three states may be observed in Fig.~\ref{fig:Identical_states}.
We note that similar states were previously numerically studied \cite{elana}, albeit with a different naming scheme.% to describe the dynamics of their model. 

%For the parameter ranges described in Sec.~\ref{sect:param}, 
We also note that one can observe states beyond (i-iii).  For example, we have observed systems for which the oscillators are incoherent regardless of whether the switches are all on or all off (e.g., for small $K$) or the switches remain on regardless of whether or not the oscillators synchronize (e.g., when $K^{\theta}$ is very small). Therefore, under the assumption that $K,K^{\theta}>0$, the I-On and S-Off states essentially decouple the oscillators and switches, leaving the existing framework for the Kuramoto and Hopfield models sufficient to capture their dynamics. We also note that states (i-iii) may not be exhaustive in other parameter regimes and network topologies, which should be the subject of future studies.

%%%%%%%%%%%%%%%%%%%%%%%%%%%%%%%%%%%%%%%%%%%%%%%%%%%%%%%%
\section{Analysis}\label{sect:analysis}
%%%%%%%%%%%%%%%%%%%%%%%%%%%%%%%%%%%%%%%%%%%%%%%%%%%%%%%%

We now analyze the three steady state solutions of interest. In Sect.~\ref{sect:analysis_IO} and Sect.~\ref{sect:analysis_SO} we respectively study solutions for the I-Off and S-On states. In Sect. \ref{sect:analysis_SP} we study the S-P state for two phase lag distributions: identical phase lags (Sect. \ref{sect:analysis_SP_i}) and uniformly-distributed phase lags (Sect.~\ref{sect:analysis_SP_u}), which respectively represent the limiting cases of very homogeneous and very heterogeneous switches. While the analyses in Sect.~\ref{sect:analysis_IO} and Sect.~\ref{sect:analysis_SO} only assume large system size, the analysis presented in Sect.~\ref{sect:analysis_SP} additionally assumes that coupling adaptation is slow compared to the switch and oscillator dynamics, $\tau\gg \max\{1, \omega_0 ^{-1}\}$. The relaxation of assumptions made in Sec.~\ref{sect:analysis_SP} is addressed in Sect.~\ref{sect:num_relax}.

%%%%%%%%%%%%%%%%%%%%%%%%%%%%%%%%%%%%%%%%%%%%%%%%%%%%%%%%
\subsection{The Incoherent-Off state}\label{sect:analysis_IO}
%%%%%%%%%%%%%%%%%%%%%%%%%%%%%%%%%%%%%%%%%%%%%%%%%%%%%%%%

We first consider the I-Off steady state solution, which is the equilibrium solution of Eqs.~(\ref{eq:system1}-\ref{eq:system3}) in which $x_m=-\eta$ and $\tilde x_m=0$ for all $m$, $r_x=0$, $r_\theta=0$, and $k^{}=0$. Note that we assume $\eta>0$ and $N\to\infty$. In this solution, oscillators evolve independently of each other and their phases are given by $\theta_n(t) = \omega_nt+\theta_n(0)$. 

In Fig.~\ref{fig:Identical_states}(a) we show a simulation that approaches this steady state solution, where a system with $N=M=1000$ oscillators and switches is initialized with $k^{}(0)=4$, random values $\{\theta_n\}$ chosen such that $r_\theta(0)\approx0.6$, and random values $\{x_m\}$ such that $r_x(0)\approx0$ and the set $\{x_m(0)\}$ centered at -1. As time increases, $r_x$ remains at 0 for all time $t$, each $x_m$ decays to $-\eta$, and both $r_\theta$ and $k^{}$ decay to $0$. From Eqs.~(\ref{eq:system3}), one can see that the decay of $k^{}$ is described by $k^{}(t)=k^{}(0)e^{-t/\tau}$ since $r_x=0$.

%%%%%%%%%%%%%%%%%%%%%%%%%%%%%%%%%%%%%%%%%%%%%%%%%%%%%%%%
\subsection{The Synchronized-On state}\label{sect:analysis_SO}
%%%%%%%%%%%%%%%%%%%%%%%%%%%%%%%%%%%%%%%%%%%%%%%%%%%%%%%%

We next consider the S-On state in which the oscillators remain synchronized ($r_\theta>0$) and all switches remain on ($r_x=1$). Assuming $r_x=1$ and looking for an equilibrium of Eqs.~(\ref{eq:system1}-\ref{eq:system3}) for large $N$ and $M$, we first note that Eq.~(\ref{eq:system3}) implies $k^{}=K$. Using this fixed value for $k^{}$, we examine the synchronization of oscillators under fixed coupling strength. Assuming that the frequency distribution $\Omega(\omega)$ is unimodal, smooth, and symmetric about its mean $\omega_0$, the order parameter $r_\theta$ is given implicitly for $k^{}>K_0 \equiv \pi^{-1}2/\Omega(0)$ by the nonzero solution of the equation \cite{Kuramoto1}
\begin{equation}
%R(r) = r - k^{\theta}r\int_{-\pi/2}^{\pi/2} \cos^2\Omega(k^{\theta}r\sin\theta)d\theta(r) 
1 = k^{}\int_{-\pi/2}^{\pi/2} \cos^2\Omega(k^{}r_\theta\sin\theta)d\theta .
\label{eq:consist0}
\end{equation}
Here $K_0$ is referred to as the critical coupling strength as the oscillators will deterministically attain the incoherent state whenever $k\le K_0$ [which is always the case for the proposed hybrid model, Eqs.~(\ref{eq:system1}-\ref{eq:system3}), when $K\le K_0$].
While one can numerically solve the above to determine the dependency of $r_\theta$ on $k^{\theta}$ for arbitrary distributions $\Omega(\omega)$, it may be directly integrated for a Lorentzian distribution $\Omega(\omega)={\pi^{-1} \Delta}/{[(\omega-\omega_0)^2+\Delta^2]}$ yielding 
\begin{equation}
r_\theta = \left\{ \begin{array}{ccc} 0&,&k^{}<K_0\\ \sqrt{1-\frac{K_0 }{k^{}}}&,&k^{}\ge K_0 ,\end{array} \right. \label{eq:r_L}
\end{equation}
where $\Delta$ represents the spread in frequencies and $K_0=2\Delta$. When oscillators synchronize, they rotate together with a mean field phase $\psi(t) = \omega_0t +\psi(0)$, where $\psi(0)$ depends on initial conditions.

Having described the macroscopic dynamics of the S-On state, we now turn to the internal switch dynamics $x_m$ for this solution (recall that the external switch states are given by $\tilde x_m=1$ for all $m$ to be consistent with $r_x=1$). Using that $\psi(t) = \omega_0t  +\psi(0)$ and that both $r_x$ and $r_\theta$ are fixed, we directly integrate Eq.~(\ref{eq:system1}) to find
\begin{eqnarray}
x_m(t) &=& \hat x_m(t) -  e^{-(t-t_0)}D_m^{}\label{eq:x_ident_0},
\end{eqnarray}
where $D_m$ is a constant that depends on initial conditions,
\begin{eqnarray}
\hat x_m(t) &=& A^{} +  C^{}  \sin(\omega_0 t -\delta-\beta_m) \label{eq:x_ident}
\end{eqnarray}
is the steady state solution, and we have defined the following constants,
\begin{align}
A^{}&=\Big( K^{x} - \eta \Big),  \\
C^{}&=K^{\theta}r_\theta\cos(\delta), \\
%D_m^{(1)}&=A^{(1)} +  C^{(1)}  \sin(\omega_0 t_0 -\delta-\beta_m) -x_m(t_0),  \\
\delta &= \arccos \left( 1/\sqrt{1+\omega_0^2} \right).\label{eq:delta}
\end{align}
In the limit $t\to\infty$, the second term in Eq.~\eqref{eq:x_ident_0} decays, and thus all internal switch variables approach similar trajectories described by Eq.~\eqref{eq:x_ident}. Specifically, they attain oscillatory trajectories with a mean value $A^{}$ and an oscillation amplitude $C^{}$.

 One prediction of this result is that to be self-consistent with our definition of the S-On state (i.e., $r_x=1$ for all $t$), we require that $\tilde x_m=1$ and $x_m>0$ for all $t$ and $m$. Because $x_m(t)$ obtains its minimum at $A^{}-C^{}$, the existence of a S-On solution requires parameters such that $A^{}>C^{}$, implying that $K^{x}$ should be larger than a critical value $K^{x}_1$ given by
\begin{equation}
K^{x}_1= \eta  + K^{\theta}r_\theta\cos(\delta) \label{eq:k1}.
\end{equation}
For a Lorentzian frequency distribution $\Omega(\omega)$ we have
\begin{equation}
K^{x}_1= \eta  + K^{\theta}  \sqrt{ \frac{1-2\Delta/K}{1+\omega_0^2}}.
\end{equation}

Another result is that in the S-On state, the only difference between the switches' internal variables $\{x_m\}$ is the phase at which they oscillate [see Eq.~\eqref{eq:x_ident}]. It follows that for a given distribution of phases $B(\beta)$, we may predict the distribution of internal switch parameters, $\rho^{}(x)$, which may or may not depend on time. Of particular interest are the limiting cases of identical phase lags and uniformly-distributed phase lags, $B(\beta)=(2\pi)^{-1}$ for $\beta\in[-\pi,\pi]$ and 0 otherwise. For identical phase lags, $\beta_m=\beta$ for all $m$, all switches have internal variables with identical trajectories $x_m(t)= A^{} +  C^{}  \sin(\omega_0 t -\delta-\beta) $. For uniformly distributed phase lags in the asymptotic limit $M\to\infty$, the distribution of possible $x$ values for a randomly selected switch is given by
\begin{eqnarray}
\rho^{}(x)&=& \frac{1}{\pi \sqrt{C^2-(x-A^{})^2 }}.\label{eq:rho1}
\end{eqnarray}
This distribution is obtained by solving Eq.~\eqref{eq:x_ident} for $\beta_m(\hat x_m)$ and simplifying $\rho^{}(x) = \left| d  \beta_m(x) / dx \right| B(\beta_m(x))$ using that $\cos(\arcsin(s)) = \sqrt{1-s^2}$.

In Fig.~\ref{fig:Identical_states}(b) we confirm these results by showing time series for dynamics approaching the S-On state solution. The system containing $N=M=1000$ oscillators and switches is initialized with $k^{}(0)=4$, random values $\{\theta_n\}$ chosen such that $r_\theta(0)=0.7$, and random values $\{x_m\}$ such that $\langle x_m(0) \rangle=3$. For these initial conditions, $r_x$ quickly approaches and remains at $r_x=1$. Because a Lorenzian distribution of frequencies $\Omega(\omega)$ was used, $r_\theta$ converges to its expected solution $r_\theta=\sqrt{(1-2\Delta/K)}=\sqrt{2/3}$. One can also observe that $k^{}$ approaches its expected value of $k^{}=K=6$. Assuming that $r_x$ is constant, Eq.~(\ref{eq:system3}) implies that $k^{}$ converges exponentially to $K$ with time constant $\tau$.

%%%%%%%%%%%%%%%%%%%%%%%%%%%%%%%%%%%%%%%%%%%%%%%%%%%%%%%%
\subsection{The Synchronized-Periodic state}\label{sect:analysis_SP}
%%%%%%%%%%%%%%%%%%%%%%%%%%%%%%%%%%%%%%%%%%%%%%%%%%%%%%%%

We now analyze steady state solutions in which the oscillators synchronize and each switch $m$ periodically fluctuates between the on ($\tilde x_m=1$) and off ($\tilde x_m=0$) states. Our analysis assumes that both $N$ and $M$ are large and that the adaptation in coupling strength is slow compared to the oscillator and switch dynamics, $\tau\gg\max\{1,\omega_0^{-1}\}$. {This separation of timescales will allow us to simultaneously consider the steady state behavior of the dynamics of switches and oscillator phases, which evolve at the fast time scale [i.e., Eqs.~(\ref{eq:system1}-\ref{eq:system2}) while assuming that $k^{}$ is approximately constant], as well the dynamics of coupling adaptation, which evolves at the slow time scale [i.e., Eq.~(\ref{eq:system3}) while assuming the fast dynamics approximately remain in a steady state]. The relaxation of this large $\tau$ assumption is numerically studied in Sect. \ref{sect:num_relax_tau}.}

The nature of the S-P state strongly depends on the distribution of phase lags $B(\beta)$. Therefore, in this section we study the limiting cases in which either (1) the distribution of phase lags is uniform, $B(\beta)=(2\pi)^{-1}$, or (2) all the phase lags are identical, $\beta_m = \beta$ for all $m$. In Sec.~\ref{sect:num_relax_lag} we find that the results for more general unimodal phase lag distributions behave as an interpolation between the results for these two cases.

%%%%%%%%%%%%%%%%%%%%%%%%%%%%%%%%%%%%%%%%%%%%%%%%%%%%%%%%
\subsubsection{Uniformly distributed phase lags}\label{sect:analysis_SP_u}
%%%%%%%%%%%%%%%%%%%%%%%%%%%%%%%%%%%%%%%%%%%%%%%%%%%%%%%%

We now study the steady state solution for the situation in which the phase lags $\{\beta_m\}$ are uniformly distributed in $[-\pi,\pi]$ (i.e., $B(\beta)=(2\pi)^{-1}$ for $\beta\in[-\pi,\pi]$ and 0 otherwise), which is the most heterogeneous distribution of phase lags. We begin our analysis by assuming that the system is in the S-P state, the coupling strength adaptation is slow, $\tau\gg \max\{1, \omega_0^{-1} \}$, and the system size is large, $N,M\to\infty$. Motivated by our results from the previous section, we look for a solution in which $r_x$ and $r_\theta$ are time independent. 

Letting $r_x$ be constant, Eq.~(\ref{eq:system3}) has an equilibrium value of $k^{} = K r_x $. It follows that the order parameter $r_\theta$ is given by Eq.~\eqref{eq:r_L} with $k^{}=K r_x $. Note that our assumption that the oscillators synchronize further restricts our interest to values such that $K r_x>K_0$. Also, recall that the synchronized oscillators rotate with a mean field $\psi(t) = \omega_0t +\psi(0)$.
Using these explicit descriptions for $k^{}$, $r_\theta$, and $\psi$, we can again directly integrate Eq.~(\ref{eq:system1}). Neglecting the transient part of this solution [e.g., see Eqs.~(\ref{eq:x_ident_0}-\ref{eq:x_ident})], we find that the switches' internal variables follow trajectories described by
\begin{equation} 
x_m(t) = D + E\sin(\psi-\delta-\beta_m),\label{eq:x_10}
\end{equation}
where $\delta$ is defined in Eq.~(\ref{eq:delta}) and
\begin{eqnarray}
D&=&K^{x}r_x-\eta ,\\
E&=&K^{\theta}\cos(\delta) \sqrt{1-K_0/(Kr_x)} .
\end{eqnarray} 
As in the derivation of Eq.~\eqref{eq:rho1}, the distribution of $x_m$ values for a randomly selected switch is given by
\begin{eqnarray}
\rho^{}(x)&=& \frac{1}{\pi \sqrt{E^2-(x-D)^2 }}. \label{eq:rho2}
\end{eqnarray}
Because the time-invariant fraction of switches in the on state, $r_x$, corresponds to the fraction of switches with positive $x_m$, i.e., $r_x=\int_0^\infty \rho^{}(x) dx $, we can insert $\rho$ from Eq.~\eqref{eq:rho2} to write a consistency equation for $r_x$. After integration we obtain
\begin{equation}
F(r_x) \equiv r_x-1/2-\pi^{-1}\arcsin \big({D}/{E}\big) = 0, \label{eq:F2}
\end{equation}
where $r_x$ values solving $F(r_x)=0$ are potential solutions for the S-P state. Therefore, we found that when the distribution of phase lags is uniform, there is a potential solution in which each switch turns on and off periodically, but the fraction of switches that are on remains constant and can be found by solving a self-consistency condition, Eq.~\eqref{eq:F2}.

\begin{figure*}[t]
\includegraphics[width=18cm]{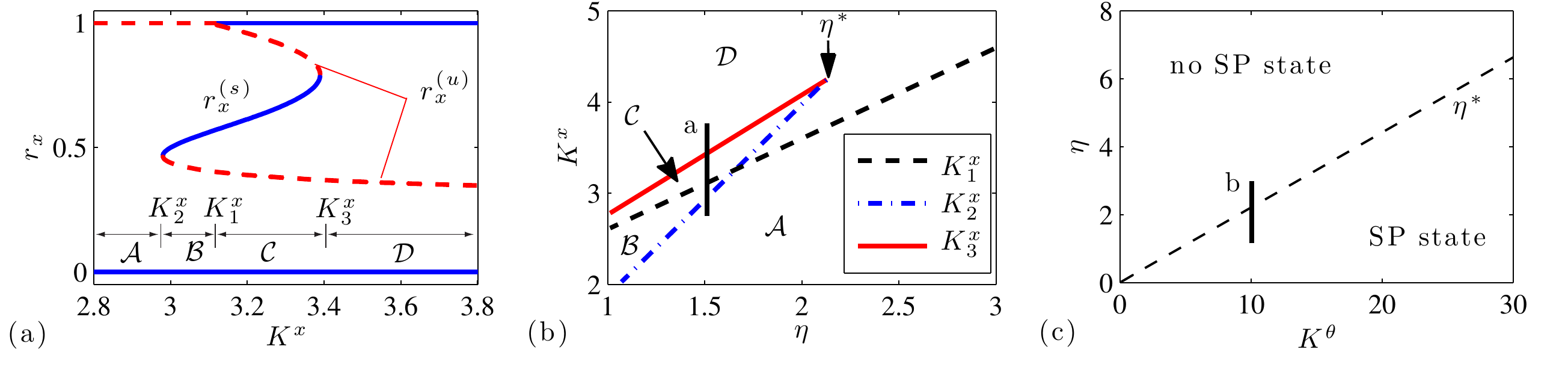}
\caption{(Color online) 
Uniformly distributed phase lags $\beta_n\in[-\pi,\pi]$.
(a) Steady state solutions for the I-Off, S-On, and S-P states are respectively shown by $r_x=0$, $r_x=1$, and solutions solving $F(r_x)=0$ for Eq.~\eqref{eq:F2}. Solid blue lines indicate stable solutions, whereas dot-dashed red lines indicate unstable solutions.
(b) Stability regions are shown for variable $K^{x}$ and $\eta$. The definitions are summarized in Table \ref{table:states}. The vertical line indicates parameter values shown in Fig. \ref{fig:bifurcation1}(a).  
Note that for $\eta>\eta^*$, the critical values $K^{x}_2$ and $K^{x}_3$ merge, corresponding to the disappearance of the stable branch $r_x^{(s)}$. 
(c) No S-P state exists for $\eta>\eta^*$. The vertical line indicates the $\eta$ and $K^{\theta}$ values shown in Fig.~\ref{fig:bifurcation1}(b).  
}
\label{fig:bifurcation1}
\end{figure*} 

In Fig. \ref{fig:bifurcation1}(a) we show numerically computed solutions of Eq.~\eqref{eq:F2}, which were determined numerically to be either stable ($r_x^{(s)}$, blue solid curved line) or unstable ($r_x^{(u)}$, red dashed curved lines). The I-Off ($r_x=0$) and S-On ($r_x=1$) states are also shown (horizontal lines). The S-On solution is only stable above the critical value $K^{x}_1$ defined by Eq~\eqref{eq:k1}. 
Due to the nature of solutions to Eq.~\eqref{eq:F2}, which gives rise to both stable and unstable branches, two additional 
critical values appear, $K^{x}_2$ and $K^{x}_3$, which respectively denote the values of $K^x$ at which the $0<r_x^{(s)}<1$ branch appears and disappears. 
%left and right reflection points at which the stable branch becomes unstable. 
These may be computed by jointly solving $F=0$ and $dF/d r_x=0$ for $(r_x,K^{x})$. 
These three critical values bound regions of $(K^x,\eta)$ phase space in which the system has similar multi-stability properties. These regions are labeled $\{\mathcal{A}, \mathcal{B}, \mathcal{C}, \mathcal{D}\}$ and their descriptions are summarized in Table~\ref{table:states}.

\begingroup
\squeezetable

\begin{table}[b]
\centering
\caption{Summary of stability regions.}
\begin{tabular}{ccc} 
     \hline     \hline
     ~region~  &~~~~& stable solutions for $r_x $ \\ \hline   \hline
  {\bf $\mathcal{A}$}  &~~~~& $   0$ \\
  {\bf $\mathcal{B}$}  &~~~~& $    r_x ^{(s)}$ and $ 0 $  \\
  {\bf $\mathcal{C}$}  &~~~~&$   1, $  $r_x   ^{(s)},$ and $ 0 $ \\
  {\bf $\mathcal{D}$}  &~~~~&$   1$ and $0 $   \\ 
  %{\bf E}  && $\langle r_x\rangle\in\{1,0\}$ are stable solutions. \\
  \hline     \hline
\end{tabular}
\label{table:states}
\end{table}
\endgroup

In Fig.~\ref{fig:bifurcation1}(b) we show the $(K^x,\eta)$ phase space depicting these stability regions for variable switch thresholds, $\eta$, and switch-switch coupling strength, $K^{x}$. The parameter values used to make Fig. \ref{fig:bifurcation1}(a) are shown by a vertical black line labeled a. Note that the critical values $K^{x}_2$ (blue dot-dashed line) and $K^{x}_3$ (red solid line) merge at a critical value $\eta^*$. For larger $\eta$ values, there is no stable branch $r_x^{(s)}$ and thus no S-P state.

In Fig. \ref{fig:bifurcation1}(c) we plot the critical threshold value $\eta^*$ as a function of the switch-oscillator coupling strength $K^{\theta}$, which may be numerically obtained by simultaneously solving $F=0$, $dF/dr_x=0$, and $d^2F/dr_x^2=0$ for $(r_x,K^{x},\eta)$. The vertical black line labeled b indicates the $\eta$ and $K^{\theta}$ values shown in Fig.~\ref{fig:bifurcation1}(b).

In summary, we have found that for fixed $K^{\theta}$, a stable $\text{S-P}$ state only exists provided that $\eta$ is sufficiently small and $K^{x}\in(K^{x}_2,K^{x}_3)$. This sensitive interplay between parameters $K^{\theta}$, $K^{x}$, and $\eta$ may be intuitively understood by considering Eq.~(\ref{eq:system1}) and observing that $\eta$ competes with $K^{x}$ and $K^{\theta}$ in determining the dynamics of $x_m$. The parameter ranges allowing the S-P state (i.e., the union of the stability regions $\mathcal{B}$ and $\mathcal{C}$) therefore represents a regime in which no parameter dominates Eq.~(\ref{eq:system1}).

\begin{figure*}[t]
\centering
\includegraphics[width=18cm]{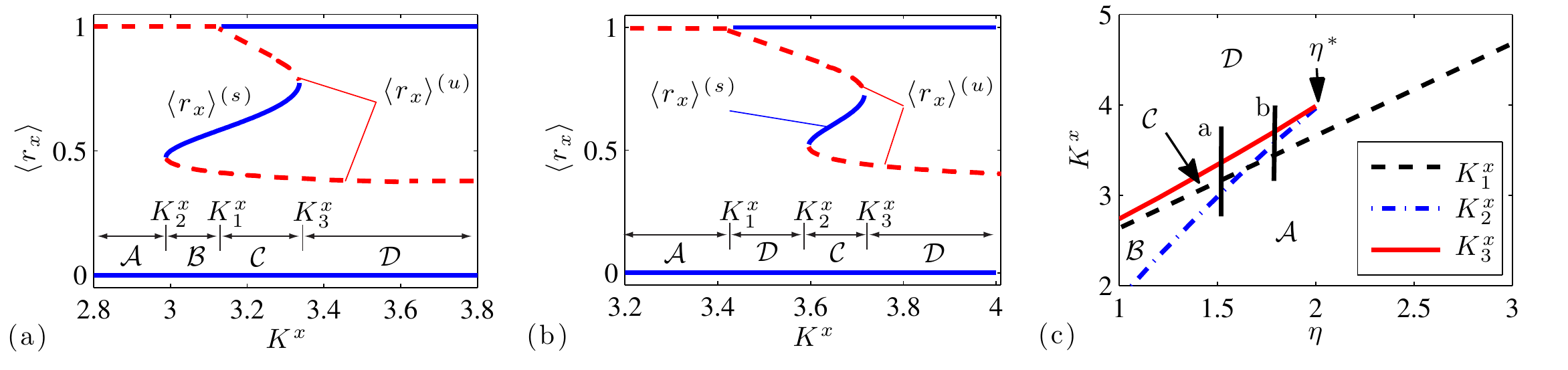}
\caption{(Color online) 
Identical phase lags. 
(a) For $\eta=1.5$, steady state solutions for the I-Off, S-On, and S-P states are respectively shown as $\langle r_x\rangle=0$, $\langle r_x\rangle=1$ , and values $\langle r_x\rangle=0$ solving the consistency equation developed in Appendix~\ref{appendix:SP_id}.
Solid (dot-dashed) lines indicate stable (unstable) solutions, where one can observe that the S-P state is only stable for $K^{x}>K^{x}_1$ given by Eq.~\eqref{eq:k1}. This critical value in addition to critical values $K^{x}_2$ and $K^{x}_3$ leads to four regions of stability.
(b) Solutions for $\langle r_x \rangle$ are shown for $\eta=1.8$, where because $K^{x}_2 > K^{x}_1$ the stability region B has been replaced by stability region D.
(c) Bifurcation diagram where critical $K^{x}$ values are shown for variable $\eta$. Vertical lines labeled a and b respectively indicate the $\eta$ and $K^{x}$ values shown in Figs. \ref{fig:bifurcation0}(a)-\ref{fig:bifurcation0}(b).   
}
\label{fig:bifurcation0}
\end{figure*}

%%%%%%%%%%%%%%%%%%%%%%%%%%%%%%%%%%%%%%%%%%%%%%%%%%%%%%%%
\subsubsection{Identical phase lags}\label{sect:analysis_SP_i}
%%%%%%%%%%%%%%%%%%%%%%%%%%%%%%%%%%%%%%%%%%%%%%%%%%%%%%%%

We now consider the S-P state for identical phase lags by letting $B(\beta_m)=\delta(\beta_m-\beta)$, i.e., $\beta_m=\beta$ for all $m$. While our analytic approach to this system is very similar to that presented in the previous section for uniformly-distributed phase lags, the analysis is slightly more involved. Therefore, for brevity we include this derivation in Appendix~\ref{appendix:SP_id} and only summarize our results here.

Motivated by the observation that the internal switch variables $\{x_m\}$ attain identical trajectories in the S-On state for identical phase lags [e.g., see Eq.~\eqref{eq:x_ident} for $\beta_m=\beta$], it is expected and observed that switches also attain identical trajectories in the S-P state for identical phase lags. It follows that $r_x$ will periodically fluctuate between 1 (all switches on) and 0 (all switches off), attaining a time-varying trajectory sometimes characterized as a ``square wave''. Moreover, as numerically observed in this and previous research \cite{elana}, this trajectory is periodic with period $T_0 = 2\pi / \omega_0$ and a duty ratio that determines its time-averaged value $\langle r_x \rangle$. See Appendix~\ref{appendix:SP_id} for details.
Importantly, these dynamics occur at a timescale much faster than $\tau$ since we assumed $\tau\gg\omega_0^{-1}$. In this limit, $k(t) = k\langle r_x \rangle + \mathcal{O}(T_0/\tau)$ and therefore we treat $k(t)$ as a constant, $k=K\langle r_x \rangle$. The order parameter $r_\theta$ reaches the value given by Eq.~\eqref{eq:r_L} with $k=K\langle r_x\rangle$. Assuming that $r_\theta$ is constant and that $r_x$ alternates between 0 and 1, Eq.~\eqref{eq:system1} can be integrated to obtain the trajectory of the switches' internal variable $x_m$ in terms of the time average $\langle r_x \rangle$. Finally, a self-consistency equation is obtained by requiring that these trajectories result in the same average value $\langle r_x \rangle$.

In Fig.~\ref{fig:Identical_states}(c) we show time series for our system with identical phase lags approaching the SP state. As expected, $r_x$ periodically alternates between 0 and 1 with frequency $\omega_0$. In addition, $k^{}$ approaches its expected value $K\langle r_x\rangle$ and $r_\theta$ approaches its expected value given by Eq.~\eqref{eq:r_L} with $k=K\langle r_x\rangle$ (although slight fluctuations can be observed for both variables since $T_0/\tau$ is nonzero).

In Fig.~\ref{fig:bifurcation0} we show the value of $\langle r_x \rangle$ for S-P solutions found by our consistency equation (see Appendix~\ref{appendix:SP_id}) as a function of $K^{x}$ for $K=6$, $K^{\theta}=10$, $\eta=1.5$ [Fig. \ref{fig:bifurcation0}(a)] and $K=6$, $K^{\theta}=10$, $\eta=1.8$ [Fig. \ref{fig:bifurcation0}(b)]. As in the previous section, this consistency equation can give rise to several solutions $\langle r_x \rangle\in[0,1]$. These often include a stable solution ($\langle r_x\rangle^{(s)}$, blue curved solid line) and unstable solutions ($\langle r_x\rangle^{(u)}$, red dashed lines). In addition to solutions for the S-P state, solutions for the I-Off and S-On states are also shown (horizontal lines), which are respectively at $\langle r_x\rangle=0$ and $\langle r_x\rangle=1$. Note that the S-On state is only stable for $K^{x}>K^{x}_1$, given by Eq.~\eqref{eq:k1}.
In Figs.~\ref{fig:bifurcation0}(a) and \ref{fig:bifurcation0}(b), we indicate the ranges of $K^{x}$ that correspond to the regions described in Table \ref{table:states}. Note that because $K_2^{x}$ is larger than $K^{x}_1$ in Fig.~\ref{fig:bifurcation0}(b), the order of the regimes as $K^{x}$ is increased is different to that in Fig.~\ref{fig:bifurcation0}(a).

%In Fig.~\ref{fig:bifurcation0}(a) we show possible steady states for $K=6$, $K^{\theta}=10$, $\eta=1.5$, and variable $K^{x}$. Note that four regions of stability may be observed, $\{\mathcal{A},\mathcal{B},\mathcal{C},\mathcal{D}\}$. These are described by Table \ref{table:states} under the variable substitution $r_x\mapsto \langle r_x\rangle$. As discussed in the previous section, the locations of $K^{x}_2$ and $K^{x}_3$, which bound the stability regions, may be found using the consistency equation.
%In Fig.~\ref{fig:bifurcation0}(b), we show solutions $\langle r_x \rangle$ for $K=6$, $K^{\theta}=10$, $\eta=1.8$, and variable $K^{x}$. Observe that after increasing the switch thresholds from $\eta=1.5$ to $\eta=1.8$, the critical value $K^{x}_2$ is now larger than $K^{x}_1$, and therefore the stability region labeled $\mathcal{B}$ in Fig.~\ref{fig:bifurcation0}(a) now corresponds to stability region $\mathcal{D}$ in Fig.~\ref{fig:bifurcation0}(b). 

In Fig.~\ref{fig:bifurcation0}(c) we provide a bifurcation diagram summarizing the stability regions for variable switch-switch coupling strength, $K^{x}$, and switch thresholds, $\eta$. Note that for $\eta>\eta^*$ (the value at which $K^{x}_2$ and $K^{x}_3$ merge), there is no stable solution $\langle r_x \rangle$ and hence no S-P state. The vertical black lines labeled a and b respectively indicate the $\eta$ and $K^{x}$ values shown in Figs.~\ref{fig:bifurcation0}(a) and\ref{fig:bifurcation0}(b). 

In summary, although the temporal dynamics of the stable S-P states differ greatly for switches with uniformly distributed phase lags and identical phase lags (e.g., $r_x$ is either constant or periodically fluctuates), the underlying state space is very similar [e.g., compare Fig.~\ref{fig:bifurcation1}(b) to Fig.~\ref{fig:bifurcation0}(c)]. In both cases a stable S-P state only exists for a regime in which the parameters of the three terms describing the dynamics of the internal switch variables $\{x_m\}$ (i.e., $\eta$, $K^{x}$, and $K^{\theta}$) are chosen such that no single term dominates Eq.~\eqref{eq:system1}.
%
%Specifically, if the switch threshold $\eta$ is too large (i.e., $\eta>\eta^*$), then it dominates the dynamics and there is a single transition from region $\mathcal{A}$ to region $\mathcal{D}$ as $K^{x}$ increases [see Fig. \ref{fig:bifurcation0}(c)]. Even if $\eta$ is sufficiently small, if $K^{x}$ is either too small or too large with respect to $K^{\theta}$ [i.e., $K^{x}\not\in(K^{x}_2,K^{x}_3)$], then the relative influence of switches and oscillators is not ``tuned'' to allow the more complex S-P state.

%%%%%%%%%%%%%%%%%%%%%%%%%%%%%%%%%%%%%%%%%%%%%%%%%%%%%%%%
\section{Numerical Investigations}\label{sect:num}
%%%%%%%%%%%%%%%%%%%%%%%%%%%%%%%%%%%%%%%%%%%%%%%%%%%%%%%%

Having introduced our hybrid model, the steady states of interest, and our analysis, we now illustrate our results and numerically explore further dynamics.
In Sec.~\ref{sect:num_trans} we investigate transitions between the steady state solutions, which may either be deterministically onset by the slow variation of a parameter (Sec.~\ref{sect:num_bif}) or stochastically onset by finite-size fluctuations (Sec.~\ref{sect:num_spont}). In Sec.~\ref{sect:num_relax} we broaden the scope of our analysis by numerically studying the relaxation of the assumptions made in Sec.~\ref{sect:analysis_SP}. Specifically, in Sec.~\ref{sect:num_relax_lag} we study unimodal phase lag distributions, whereas in Sec.~\ref{sect:num_relax_tau} we relax the assumption of slow coupling adaptation. 

\subsection{Transitions between steady state solutions}\label{sect:num_trans}

Here we validate our analysis and explore two mechanisms that can cause transitions between the I-Off, S-On, and S-P steady state solutions: (1) deterministic transitions onset by the slow variation of a parameter (e.g., $K^{x}$) and (2) spontaneous transitions onset by fluctuations arising for systems of finite-size.
%
%Specifically, because our analysis has been developed for asymptotically large systems for which $N,M\to\infty$, our system, Eqs.~(\ref{eq:system1}-\ref{eq:system3}), is well described by asymptotic theory provided that $N$ and $M$ are sufficiently large. It follows that in this regime, dynamic bifurcation theory \cite{DynBif} accurately describes transitions between steady states when they are induced by slowly varying a system parameter. 
%
%On the other hand, for systems with moderate-to-small size, transitions can occur unexpectedly due to the discrepancy between asymptotic theory and the dynamics of finite-sized systems, often referred to as finite-size fluctuations. 
% typically $\mathcal{O}(N^{-1/2})$ for systems of coupled oscillators \cite{daido}.
%
%However, recent research modeling finite-size fluctuations as a stochastic process has shown that these spontaneous transitions may be studied using theory developed for stochastic systems \cite{vary_freq}. 

\begin{figure*}[t]
\centering
\includegraphics[width=18cm]{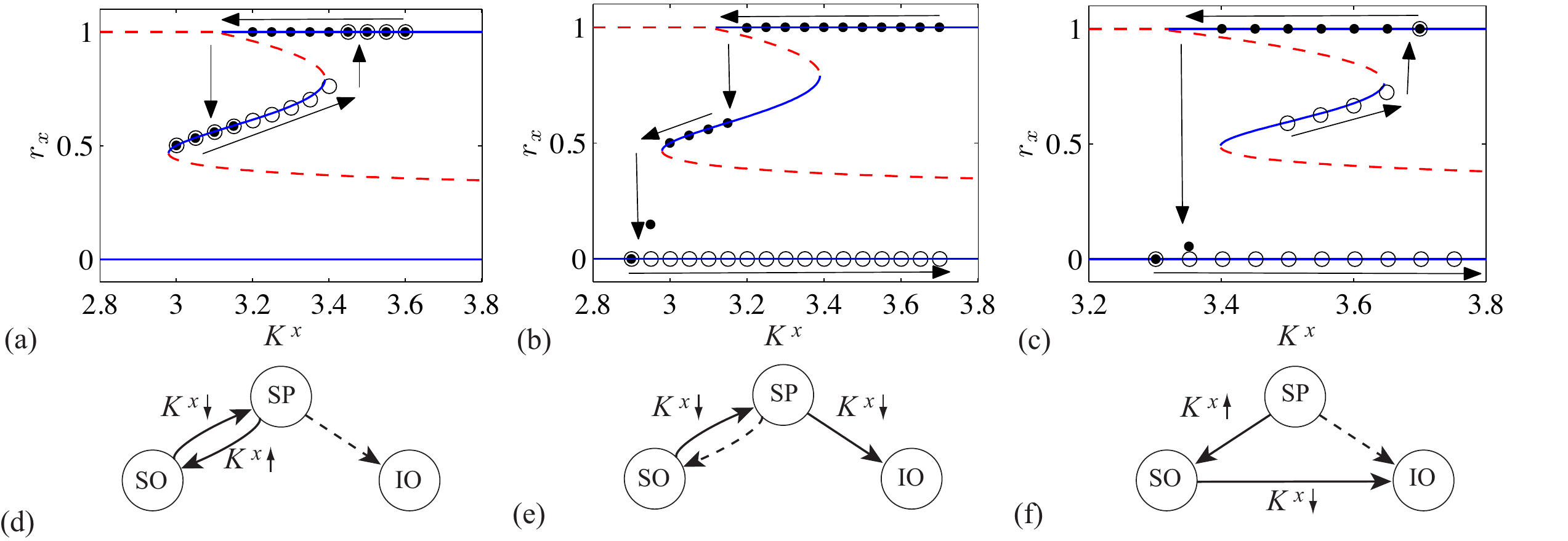} 
\caption{(Color online) Deterministic transitions onset by variable $K^{x}$.
(a-c) Trajectories of the macroscopic dynamics are shown for variable $K^{x}$. Solid and dashed lines respectively indicate stable and unstable solutions for the I-Off, S-On, or S-P states. Dots and circles respectively denote values observed for decreasing and increasing $K^{x}$. 
Three experiments are shown: (a) two reversible transitions yielding a hysteresis loop; (b) a reversible transition followed by an irreversible transition; and (c) a cascade of two irreversible transitions.
(d-e) State diagrams summarize the systems' trajectories for these three experiments. 
%Solid lines indicate dynamic bifurcations that occur while dashed lines indicate potential bifurcations that could occur under variable $K^{x}$ and all other parameters fixed. 
%State transitions are classified as reversible or irreversible based on whether or not the link between two states is bidirectional.
}
\label{fig:hist}
\end{figure*}

\subsubsection{Deterministic transitions}\label{sect:num_bif}

%In this section we numerically validate our analysis while studying deterministic transitions between the stable solutions for the IO, SO, and SP states. Specifically, 
To validate the predicted deterministic transitions between the stable solutions for the I-Off, S-On, and S-P states, we consider several simulations. In these simulations our system is initialized near a particular stable solution for given parameters $K$, $K^{\theta}$, $K^{x}$, $\tau$, and $\eta$. Then $K^{x}$ is slowly varied to explore this branch and other branches denoting stable solutions. When $K^{x}$ is varied such that the current state's solution becomes unstable, the system deterministically transitions to a solution that is stable. This method thus allows us to both confirm the accuracy of our analysis for stable solutions and study transitions onset by variable $K^{x}$ (which may be further studied as dynamic bifurcations \cite{DynBif}).
%We note that this technique avoids the difficulty of guessing initial conditions to place the system in the different basins of attractions for the different states.

%Transitions between steady state solutions that occur due to the slow variation of a control parameter (e.g., $K^{x}$) may be described by dynamic bifurcation theory \cite{DynBif}. In this section we study these expected, or deterministic, transitions by observing the state of our system for several explicit trajectories $K^{x}(t)$. 
%
%These trajectories are found to lead to discontinuous-reversible and discontinuous-irreversible transitions, where we characterize the reversibility of a transition based on whether or not a transition in the opposite direction could also occur due to variable $K^{x}$.
%We do this using the following numerical experiments with the following parameters selected for the following reason.

To allow for the numerical study of all three steady state solutions, we restrict our exploration to parameter regimes allowing for all three states (i.e., parameter regimes near stability region $\mathcal{C}$). Based on our analytical results as well as many simulations with various parameter choices, we select the following parameters for our numerical experiments: 
(i) $N=1000$ and $M=1000$ are chosen to be sufficiently large such that our asymptotic analysis for $N,M\to\infty$ is accurate. 
(ii) The frequencies $\omega_n$ are chosen from a Lorentzian $\Omega(\omega)$ with mean $\omega_0=5$ and spread $\Delta =1$. Choosing $\omega_0=\mathcal{O}(1)$ ensures that the timescales of the first and last terms in the r.h.s. of Eq.~\eqref{eq:system1} are similar.
(iii) The phase lags $\beta_m$ are chosen uniformly from $[-\pi,\pi]$ to represent a very heterogeneous system.
(iv) The parameter determining the timescale of coupling adaptation ($\tau=25$) is chosen to be sufficiently large such that a separation of timescales analysis is valid. 
(v) The maximal oscillator coupling strength $K=6$ is chosen to be sufficiently large such that the oscillators can synchronize in both the S-On and S-P states. For the parameters studied here, we found that choosing $K\ge3K_0$ typically sufficed.
(vi) The switch thresholds $\eta<\eta^*$ are chosen to be sufficiently small to allow a stable S-P state. 
(vii) The switch-oscillator and switch-switch coupling strengths ($K^{\theta}$ and $K^{x}$, respectively) are chosen such that no single term dominates Eq.~\eqref{eq:system1}. This was shown to be the case for $K^{\theta}=10$ and $K^{x}\sim 3$ in Figs.~\ref{fig:bifurcation1} and \ref{fig:bifurcation0}.
%
%Our three numerical experiments therefore consider simulations with initial conditions chosen to place the system in a specific state for these parameters and that slowly vary $K^{x}$.

In Figs.~\ref{fig:hist}(a)-\ref{fig:hist}(c) we show three such numerical experiments, each of which involves keeping all other parameters fixed while slowly varying $K^x$ at a coarse-grained rate of $dK^{x}/dt = \pm 0.005$. The trajectories shown were chosen to validate the accuracy of our results for all three steady states and to highlight the possible transitions between these states (e.g., S-P$\to$I-Off, S-P$\to$S-On, S-On$\to$S-P, and S-On$\to$I-Off). In these figures, blue solid and red dashed lines respectively indicate stable and unstable solutions, whereas filled and open circles indicate values observed from directly simulating Eqs.~(\ref{eq:system1}-\ref{eq:system3}) for decreasing and increasing $K^{x}$, respectively.
%
%, where we simulate Eqs.~(\ref{eq:system1}-\ref{eq:system3}) with $N=M=1000$, $K=6$, $K^{\theta}=10$, $\tau=25$, and uniformly-distributed phase lags, while allowing $K^{x}$ to slowly evolve at a coarse grained rate of $|dK^{x}/dt| = 0.005$. Dots and circles respectively denote observed values for decreasing and increasing $K^{x}$. Solid and dashed lines respectively indicate stable and unstable equilibria for the IO, SO, and SP states as discussed in the previous section. 

In Figs.~\ref{fig:hist}(a)-\ref{fig:hist}(b) we let $\eta=1.5$ and show two simulations: Hysteresis is shown in Fig. \ref{fig:hist}(a) for a $K^{x}$ trajectory beginning at $K^{x}=3.6$, decreasing until $K^{x}=3$, and then increasing back to $K^{x}=3.6$. Note that the system is initialized and remains in the S-On state until $K^{x}$ decreases below $K^{x}_1\approx3.12$, and then it remains in the S-P state until $K^{x}$ surpasses $K^{x}_3\approx3.38$, above which the system returns to its original state.
In Fig.~\ref{fig:hist}(b) we let $K^{x}$ decrease from $3.7$ to $2.9$ and then increase back to $3.7$. As before, while $K^{x}$ decreases the system remains in the S-On state until $K^{x}$ decreases below $K^{x}_1$, at which time it transitions to the S-P state. However, when $K^{x}$ later decreases below $K^{x}_2\approx2.95$, the system irreversibly transitions to the I-Off state. It remains in this state even as $K^{x}$ increases back to its initial value.

In Fig.~\ref{fig:hist}(c) we let $\eta=1.8$ and show a trajectory involving a cascade of two irreversible transitions: $K^{x}$ is increased from 3.5 to 3.7, then it decreases from 3.7 to 3.3, and finally it increases from 3.3 to 3.75. Under this trajectory for $K^{x}$, the system undergoes the following discontinuous transitions: it is initialized and remains in the S-P state until $K^{x}$ surpasses $K^{x}_3$, then it transitions to the S-On state where it remains until $K^{x}$ decreases below $K^{x}_1$, after which it transitions to and remains in the I-Off state. 

In Figs.~\ref{fig:hist}(d-f) state diagrams summarize the three experiments shown in Figs.~\ref{fig:hist}(a-c). Solid lines indicate transitions that occur in the experiments shown in Figs.~\ref{fig:hist}(a-c), whereas dashed lines indicate potential transitions that can occur under variable $K^{x}$ (with all other parameters fixed). Reversible transitions are indicated by bidirectional links. 
%
%In addition to allowing for the summary and classification of state transitions due to dynamic bifurcations, the network topology of such state diagrams also offers important insight toward the system's current and future states. In particular, the topology of these diagrams may be useful for understanding control mechanisms for the macroscopic dynamics of more complex systems. For example, when $\eta$ was shifted from 1.5 to 1.8 for the simulations shown in Fig.~\ref{fig:hist}, the state diagram changed from a topology allowing a hysteresis loop to a topology only allowing for a cascade of irreversible transitions. Uncovering the mechanisms that induce such topological transitions in state diagrams may be useful for furthering our understanding of natural phenomena involving cascades of irreversible transitions (e.g., the cell cycle \cite{Novak}).

%
%
%paragraph 1:
% describe the problem : validate stable solutions. 
%  numerical techniques used to address it, 
%motivate parameters chosen

%pargraph 3: compare across the conditions
%paragraph 4: the transitions -- Needs to be replaced with a detailed description of the state diagrams 5(d-f) and how that addresses the question we had put on the start.

\subsubsection{Spontaneous transitions}  \label{sect:num_spont}

\begin{figure}[t]
\centering
\includegraphics[width=8.5cm]{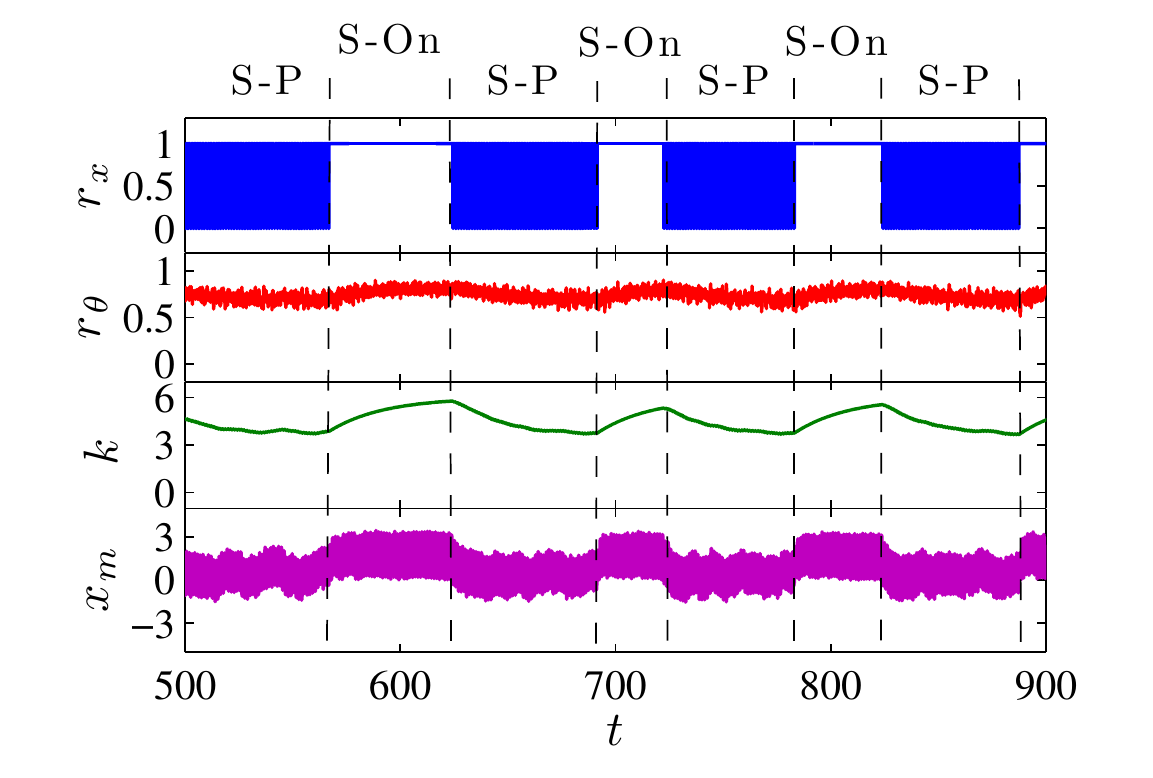} 
\caption{(Color online) Spontaneous transitions onset by finite-size fluctuations are shown between the S-On and S-P states for Eqs.~(\ref{eq:system1}-\ref{eq:system3}) with identical phase lags. These transitions are akin to the ``flickering'' phenomenon observed for stochastic processes.
% with the following parameter values: $k^{x}=3.2$, $K=6$, $k^{x\theta}=10$, $\eta=1.5$, $\beta=0$, $\tau=25$, and $N=M=100$. 
}
\label{fig:spont}
\end{figure}

In addition to transitions arising from slow change in $K^x$, transitions may also arise spontaneously due to finite-size fluctuations.  These fluctuations have been observed in the hybrid network model of Ref.~\cite{elana} and have been characterized as typically $\mathcal{O}(N^{-1/2})$ for systems of Kuramoto oscillators \cite{daido}. 
%
%In contrast to the previous section where we investigate deterministic transitions for large systems, we now study spontaneous transitions arising for smaller systems. We point out that our classification of a system size as large or small mainly depends on the time-scale of spontaneous transitions which can depend on a variety of factors, including the system parameters (e.g., $K$, $K^{x}$, and $K^{\theta}$) as well as their rates of change. For example, spontaneous transitions are also expected to occur for large systems except that the timescale required to observe transitions is too large for computer simulation. Further investigation of this phenomena is expected to be a fruitful area of future research.
%
Because we numerically observe that these finite-size effects have the most pronounced influence for our system when the switches have identical phase lags, $\beta_m=\beta$ for all $m$, in this section we focus on spontaneous transitions arising for small systems with identical switches [see Fig.~\ref{fig:spont}(b)]. 
%{\bf In particular, even after moderate search in parameter space stochastic transitions were not experimentally observed for our system with uniformly distributed phase lags.}

We now examine a state in which our system spontaneously transitions back and forth between the S-On and S-P solutions, a phenomenon which has been referred to as ``flickering'' for stochastic systems near critical transitions \cite{flicker}. 
To observe this phenomenon, we will again choose parameters such that all three states may be observed. Therefore we choose $K=6$, $K^x=3.2$, $K^{\theta}=10$, $\eta=1.5$, $\beta=0$, and $\tau=25$, placing the system in stability regime $\mathcal{C}$ [see Fig.~\ref{fig:bifurcation0}(a)]. For these fixed parameters, Eqs.~(\ref{eq:system1}-\ref{eq:system3}) were simulated for various system sizes with $N=M$ and initial conditions placing the system in the S-On state. 
In Fig.~\ref{fig:spont}(a) we plot time series for $r_x(t)$, $r_\theta(t)$, $k(t)$, and $x_m(t)$ for a simulation with $N=M=100$, where one can observe flickering between the S-On and S-P state solutions. 
% These parameter choices place the system in stability region C for which the IO, SO, and SP states all have stable solutions. The simulation is initialized in the SO state by starting all switches on and the oscillators at initial phases uniformly distributed in $[0,1]$. Note that as the system transitions between the basins of attractions for the SO and SP states, the systems' variables approach their predicted values described by our asymptotic theory (e.g., exponential decay of $k$ either to 6 or 3 for the SO and SP states, respectively).
%
As previously mentioned, this flickering phenomenon occurs due to finite-size fluctuations that spontaneously drive the system back and forth between stable equilibria (see Ref.~\cite{vary_freq}). 
We finally point out that this flickering phenomena was observed in numerical experiments by simulating with small system size (e.g., $N=M<100$) and choosing parameters placing the system near a bifurcation. Interestingly, while flickering was easily observed for our system with identical phase lags for a variety of parameter ranges, flickering has yet to be observed for uniformly distributed phase lags even after a thorough exploration of parameter space. This observation suggests that phase lag heterogeneity can significantly counter the de-stabilizing effects of finite-size fluctuations.

%In Fig. 6(b)É.. {\bf  show how rate of spontaneous transitions for both identical and heterogeneous scales with N in a figure. this would summarize the section and support our claim that heterogeneous is more stable.}

%As previously mentioned, this flickering phenomena occurs due to finite-size fluctuations that spontaneously drive the system back and forth between stable equilibria (see Fig.~4 in Ref.\cite{vary_freq}). 
%In fact a similar flickering state has been observed for systems of coupled oscillators with adaptive frequencies \cite{vary_freq}, where it was found that transitions may be described by an adaptation of Kramer's escape time formula \cite{Kramer} using that the finite size fluctuations have amplitude that is $\mathcal{O}(N^{-1/2})$ \cite{daido}. These spontaneous synchronization events were also recently utilized in systems of coupled oscillators with link adaptation to produce complex macroscopic phenomena. For example, the macroscopic dynamics of oscillator populations were shown to produce spontaneous synchronization events reminiscent of neuronal spiking \cite{macro}. 

%%%%%%%%%%%%%%%%%%%%%%%%%%%%%%%%%%%%%%%%%%%%%%%%%%%%%%%%
\subsection{Relaxing assumptions} \label{sect:num_relax}
%%%%%%%%%%%%%%%%%%%%%%%%%%%%%%%%%%%%%%%%%%%%%%%%%%%%%%%%

We have analytically studied the S-On and I-Off states for general parameter choices, as well as the S-P state for large $\tau$ and phase lags that are either identical or uniformly-distributed. We now show that our analysis also qualitatively predicts the system's dynamics for unimodal phase-lag distributions (Sect.~\ref{sect:num_relax_lag}) and for moderate-to-small $\tau$ (Sect.~\ref{sect:num_relax_tau}).

%%%%%%%%%%%%%%%%%%%%%%%%%%%%%%%%%%%%%%%%%%%
\subsubsection{Unimodal phase lags} \label{sect:num_relax_lag}
%%%%%%%%%%%%%%%%%%%%%%%%%%%%%%%%%%%%%%%%%%%

When considering unimodal phase lag distributions $B(\beta)$, the results presented in Sect.~\ref{sect:analysis_SP_u} and Sect.~\ref{sect:analysis_SP_i} respectively represent analyses of the S-P state solution for the limiting case scenarios in which $B(\beta)$ is very homogeneous or very heterogeneous. For example, if $B(\beta)$ is a normal distribution with mean $\overline \beta$ and variance $\sigma_\beta^2$, then the previous analyses represent analytic results for the limits $\sigma_\beta\to0$ and $\sigma_\beta\to\infty$. We hypothesize that if $B(\beta)$ is unimodal, then the S-P state solution can be qualitatively described by an interpolation between these two limiting cases. For example, if we vary $\sigma_\beta$ from 0 to $\infty$, we expect the trajectory $r_x(t)$ and its time averaged value $\langle r_x \rangle$ to smoothly evolve from the analytic prediction for $\sigma_\beta=\infty$ (Sec.~\ref{sect:analysis_SP_u}) to the analytic prediction for $\sigma_\beta=0$ (Sec. \ref{sect:analysis_SP_i}).

\begin{figure}[t]
\centering
\includegraphics[width=8.5cm]{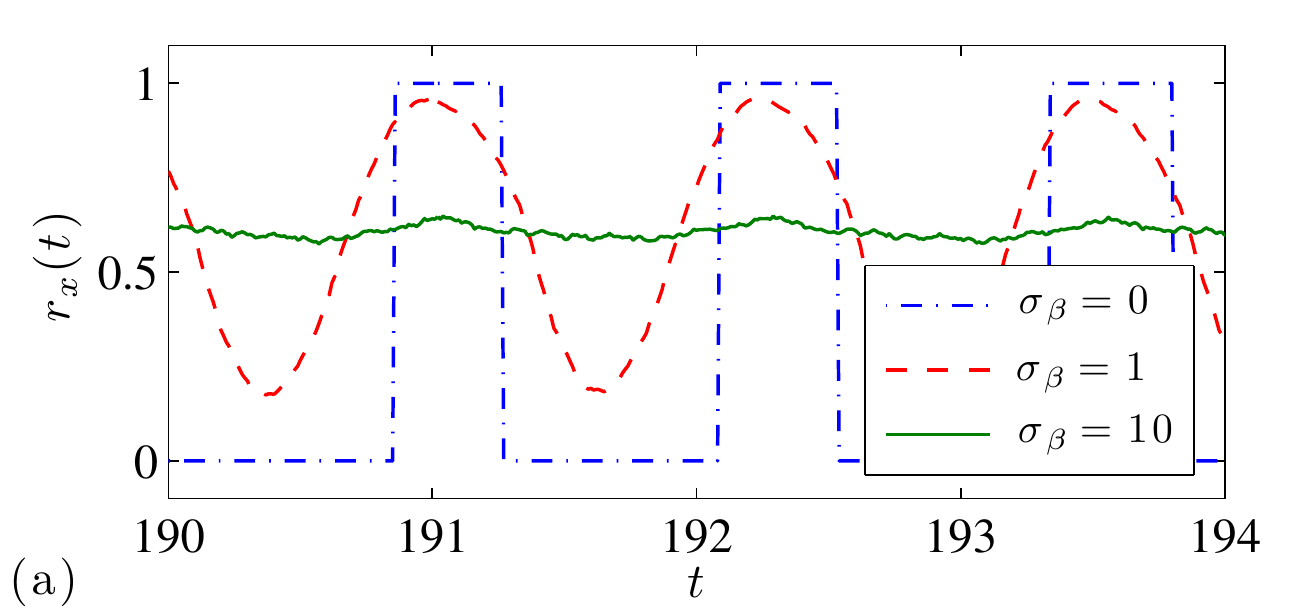} \\
\includegraphics[width=8.5cm]{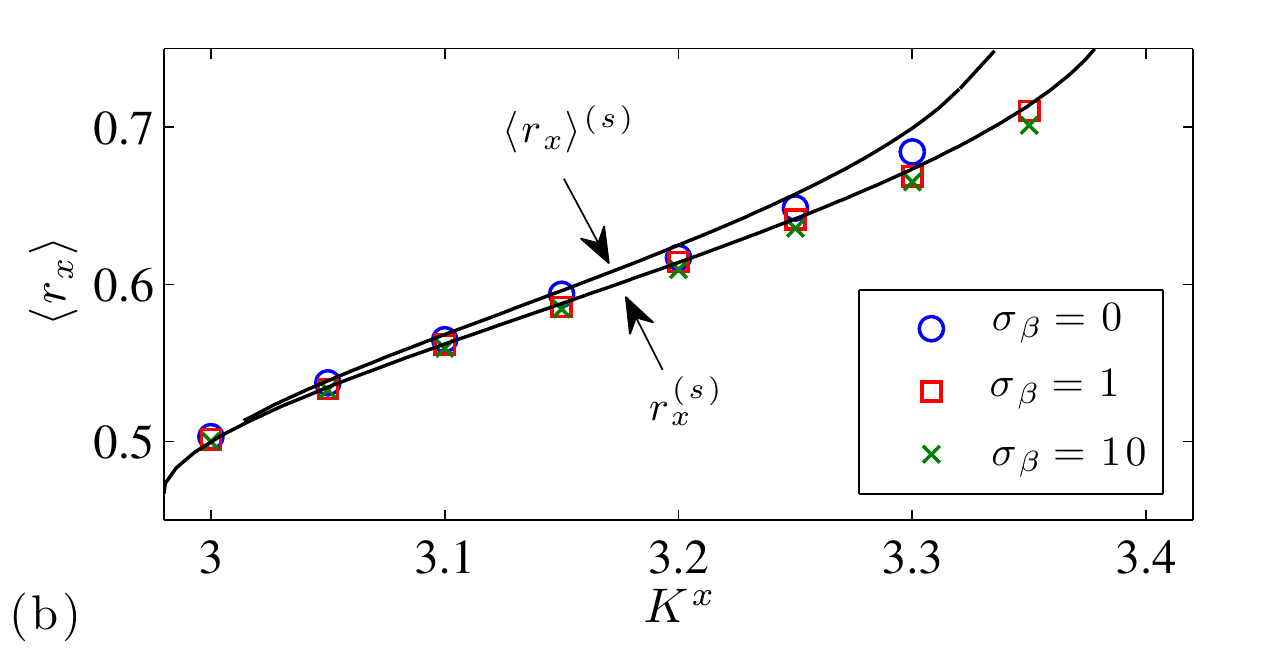} 
\caption{(Color online) 
(a) Trajectories $r_x(t)$ are shown for the S-P state with phase lags with increasing heterogeneity, $\sigma_\beta\in\{0,1,10\}$. While $r_x(t)$ is piecewise-constant for $\sigma_\beta=0$, it becomes oscillatory with decreasing amplitude as $\sigma_\beta$ increases.
(b) The underlying phase space varies only slightly for variable phase lag heterogeneity. Observed values (symbols) for the time-averaged behavior $\langle r_x\rangle$ are shown for three values of $\sigma_\beta$, which are expected to fall within the region bounded by our theoretical predictions $\langle r_x\rangle^{(s)}$ and $r_x^{(s)}$.
}
\label{fig:beta_dist1}
\end{figure} 

This conjecture is illustrated for a choice of parameters in Fig.~\ref{fig:beta_dist1} where we plot observed $r_x(t)$ trajectories [Fig.~\ref{fig:beta_dist1}(a)] and the time-average $\langle r_x \rangle $ as a function of $K^{x}$ for $\sigma_\beta\in\{0,1,10\}$ [Fig.~\ref{fig:beta_dist1}(b)]. In this figure, we plot the results from simulations with $K^{\theta}=10$, $K=6$, $K^{x}=3.2$, $\eta=1.5$, $\Delta=1$, $\omega_0=5$, and $N=M=2000$. As in previous experiments, these parameters were chosen to place the system in a regime allowing for the S-P state (i.e., stability region $\mathcal{C}$).  The system size was also chosen to be sufficiently large (i.e., $N=M=2000$) for our asymptotic analysis to be valid.

In Fig.~\ref{fig:beta_dist1}(a) one can observe that while $r_x(t)$ is a piecewise-constant periodic trajectory for $\sigma_\beta=0$, as $\sigma_\beta$ increases this trajectory becomes oscillatory with an amplitude that decays to 0 as $\sigma_\beta \to\infty$. 
%
%Thus, intermediate numbers of oscillating switches are allowed over the constant fraction in XXX assumption (Sect XXX) and the synchronized oscillations of all switches in the XXX assumption (Sect XXX).
%Experimental trajectories are shown for $K^{\theta}=10$, $K=6$, $K^{x}=3.2$, $\eta=1.5$, $\Delta=1$, $\omega_0=5$, and $N=M=2000$.
%
In Fig.~\ref{fig:beta_dist1}(b) we show the $\left( K^{x},\langle r_x \rangle \right)$ phase space, where the solid lines indicate our analytic predictions for $\sigma_\beta=\infty$ ($r_x^{(s)}$, as discussed in \ref{sect:analysis_SP_u}) and $\sigma_\beta=0$ ($\langle r_x\rangle^{(s)}$, as discussed in \ref{sect:analysis_SP_i}). As expected, numerically observed values for $\langle r_x\rangle$ with $\sigma_\beta\in\{0,1,10\}$ (symbols) are found to be near the region bounded by the two curves. It follows that although heterogeneity in phase lags has a drastic affect on the particular time-varying function that describes $r_x(t)$ for the S-P state, its average value $\langle r_x\rangle$ and the underlying phase space is only slightly affected.
%
%We therefore hypothesize that the analysis developed in Secs.~\ref{sect:analysis_SP_u} and \ref{sect:analysis_SP_i} qualitatively describes the S-P state for Eqs.~(\ref{eq:system1}-\ref{eq:system3}) for all unimodal distributions $B(\beta)$. 

%%%%%%%%%%%%%%%%%%%%%%%%%%%%%%%%%%%%%%%%%%%
\subsubsection{Moderate-to-small $\tau$} \label{sect:num_relax_tau}
%%%%%%%%%%%%%%%%%%%%%%%%%%%%%%%%%%%%%%%%%%%

The analysis presented in Sec.~\ref{sect:analysis_SP} for the S-P state assumed slow coupling adaptation, $\tau\gg \max\{1,\omega_0^{-1}\}$, and only studied steady-state solutions. We now show that recent results for the transient behavior of $r_\theta$ may be used to reduce the dimensionality of Eqs.~(\ref{eq:system1}-\ref{eq:system3}) without requiring this assumption. Specifically, it has been shown for systems of all-to-all coupled Kuramoto oscillators that the long-time dynamics of the order parameter $r_\theta e^{i\psi}$ in the asymptotic limit $N\to\infty$ is given by \cite{OA}
\begin{eqnarray}
\dot r_\theta  &=& -\Delta r_\theta + \frac{k^{}}{2} r_\theta\left(1-r_\theta^2\right),  \label{eq:OAA} \\
\dot \psi  &=& \omega_0 .\label{eq:psi}
\end{eqnarray}
We note that although this result assumes a Lorentzian frequency distribution $\Omega(\omega)$, a similar, yet more complicated, expression may be obtained and treated numerically for more general frequency distributions. We further note that it has been recently shown that these results hold even when $k^{}$ and $\Delta$ are allowed to vary with time \cite{vary_K}. 
%, and in fact this low-dimensional representation may be used to study systems that co-evolve \cite{macro} [i.e., $r_\theta$ evolves depending on $(\Delta,K^{x})$ and $\Delta$ and $K^{x}$ simultaneously evolve depending on $r_\theta$].
%, which is required by our system, 
%
%Eqs.~(\ref{eq:system1}-\ref{eq:system3}). 
Therefore, restricting our attention to the example of a Lorentzian distribution of frequencies, we find that for identical switches in the asymptotic limit $N\to\infty$, the dynamics of the S-P state is given by a system of four ordinary differential equations: Eqs.~(\ref{eq:OAA}-\ref{eq:psi}) along with
\begin{eqnarray}
\dot x  &=& -x-\eta + K^{x} \tilde x + K^{\theta} r_\theta\sin(\psi- \beta),\\
\tau \dot k^{} &=& - k^{} + K\tilde x  .\label{eq:reduced}
\end{eqnarray}
Here we have assumed that identical switches have attained identical trajectories with $x_m=x$ and $\tilde x_m=\tilde x = r_x$ for all $m$. Remarkably, the macroscopic dynamics of our $(N+M+1)$-dimensional system given by Eqs.~(\ref{eq:system1}-\ref{eq:system3}) is completely described by a three-dimensional system as $N,M\to\infty$ (since $\psi$ may be integrated).

\begin{figure}[t]
\centering
\includegraphics[width=8.5cm]{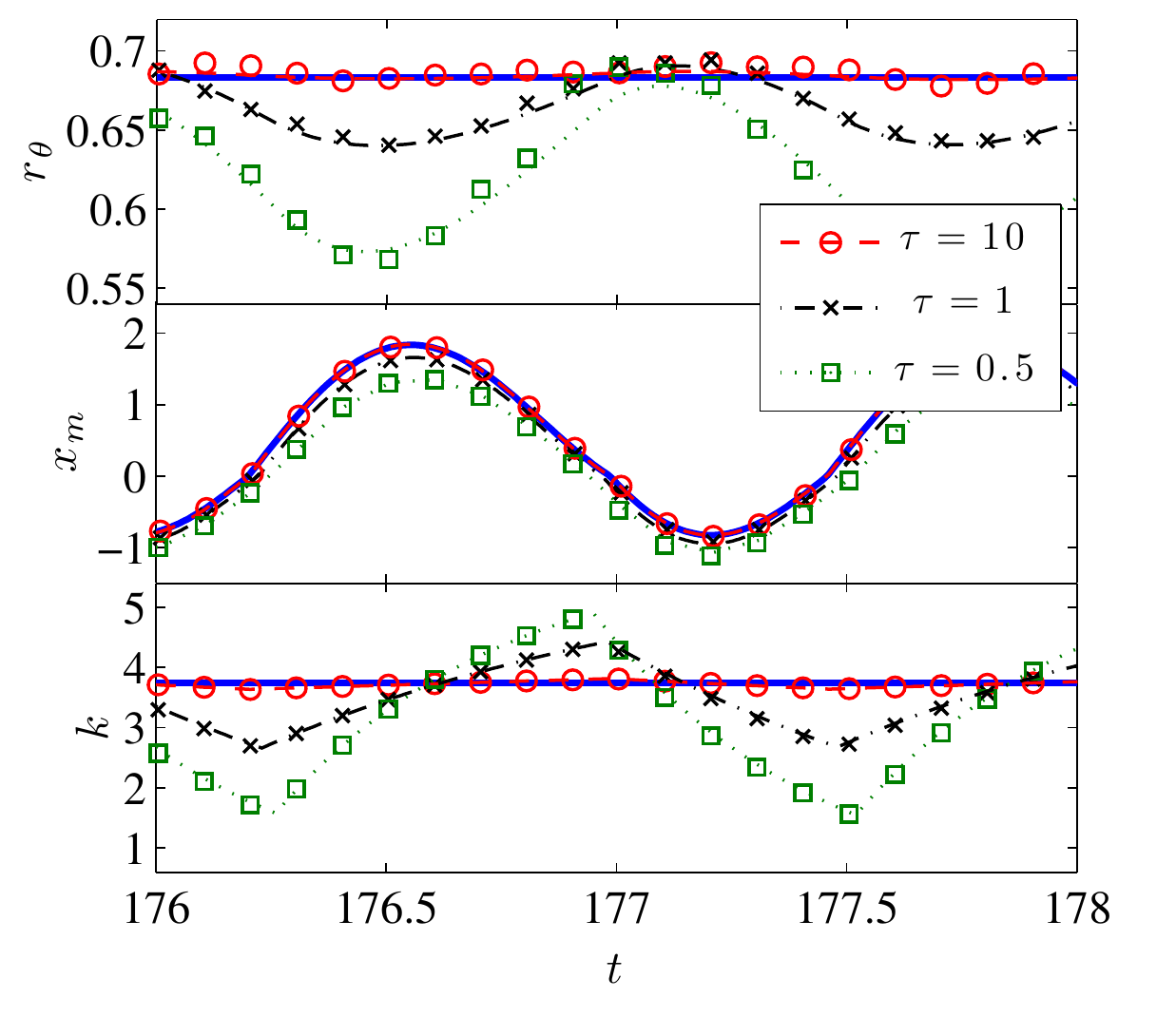} \\
\caption{(Color online) Time series are shown for identical switches in the S-P state for decreasing $\tau$. 
While our asymptotic theory accurately (thick blue lines) describes the dynamics for $\tau=10$, for small-to-moderate $\tau$ the dynamics of the high-dimensional system, Eqs.~(\ref{eq:system1}-\ref{eq:system3}) (symbols), is accurately given by the low-dimensional system, Eqs.~(\ref{eq:OAA}-\ref{eq:reduced}) (lines).
}
\label{fig:identity2}
\end{figure}

In Fig.~\ref{fig:identity2} we show that Eqs.~(\ref{eq:OAA}-\ref{eq:reduced}) (lines) accurately describe the macroscopic dynamics of the high-dimensional system Eqs.~(\ref{eq:system1}-\ref{eq:system3}) (symbols) in the S-P state. In the top, center, and lower panels we respectively plot time series for $r_\theta(t)$, $x(t)$, and $k^{}(t)$, where data is provided for three values of $\tau$. Time series are shown for times $t\in[176,178]$, which allowed enough time for the systems to approximately reach the stable S-P state. 
 %In all three panels, lines indicate simulation of the low-dimensional system given by Eqs.~(\ref{eq:OAA}-\ref{eq:reduced}), which are in good agreement with direct simulations of the high-dimensional system, Eqs.~(\ref{eq:system1}-\ref{eq:system3}) (symbols). 
 %
% note, one can observe for $r_\theta$ in the top panel that a slight discrepancy exists between Eqs.~(\ref{eq:system1}-\ref{eq:system3}) and Eqs.~(\ref{eq:OAA}-\ref{eq:reduced}), which results from finite-size fluctuations \cite{Daido,dane,macro}. 
 %
 %We note that for the high-dimensional system, $x(t)$ in the center panel reflects the trajectories $x_n(t)$ averaged over switches. 
 Initial conditions for these simulations were chosen to place the system in the basin of attraction of the S-P state by letting $r_\theta\approx0.7$, $k^{}=4$, and either $x_m$ uniformly distributed in $[-1,0]$ for Eqs.~(\ref{eq:system1}-\ref{eq:system3}) or $x=-1$ for Eqs.~(\ref{eq:OAA}-\ref{eq:reduced}). 
Other parameter values included $K^{\theta}=10$, $K=6$, $\eta=1.5$, $\Delta=1$, $\omega_0=5$, $\beta=0$, and $N=M=10^3$.
In all three panels, the thick solid blue lines indicate the predicted values using our separation of timescales analysis discussed in Sect.~\ref{sect:analysis_SP_i}. These are in good agreement with observed values for $\tau=10$. 
Interestingly, while the $r_\theta(t)$ and $k^{}(t)$ trajectories begin to fluctuate significantly as $\tau$ decreases, the $x(t)$ trajectories differ only slightly. 
%This supports our previous observation for Fig.~\ref{fig:identity3} in which we found that even though the separation of timescales assumption was violated, our theory for large $\tau$ can still predict the system's qualitative behavior. 
%
%We find that such differences disappear as $N\to \infty$, which is consistent with previous research for finite-size fluctuations in systems of Kuramoto phase oscillators with fixed $k^{\theta}$ \cite{Daido,dane,macro}. Such fluctuations will be further discussed in Sect. IVB.   

\begin{figure}[t]
\centering
\includegraphics[width=8.5cm]{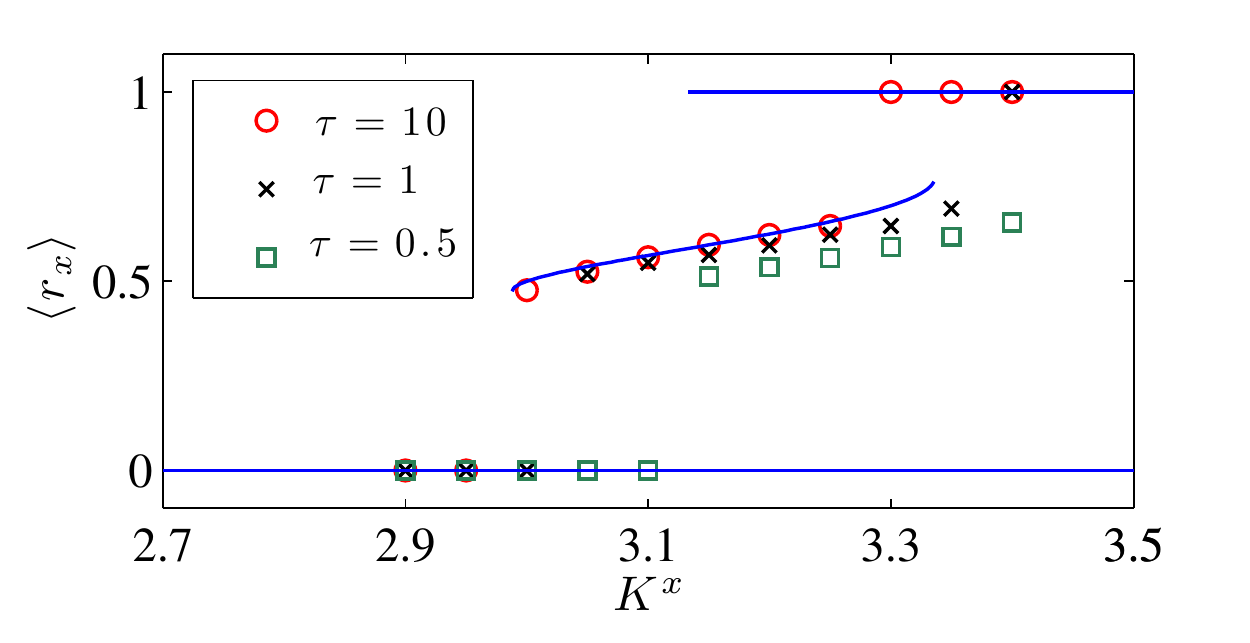} 
\caption{(Color online) 
Observed values for $\langle r_x\rangle$ from direct simulation of Eqs.~(\ref{eq:system1}-\ref{eq:system3}) (symbols) agree well with predicted values from the separation of timescales analysis for large $\tau$ (curved line). Results are shown with variable $K^{x}$ for three values of $\tau$. Note that as $\tau$ becomes small, the observed $\langle r_x\rangle$ values shift slightly to the right.
}
\label{fig:identity3}
\end{figure} 

In Fig.~\ref{fig:identity3} we show that this slight variation in $x(t)$ for decreasing $\tau$ can result in the system having a qualitatively similar phase-space if $\tau$ is not too small. Here we plot observed values of $\langle r_x \rangle$ (symbols) for the S-P state versus $K^{x}$ for several values of $\tau$. Whereas our separation of timescales analysis discussed in Sect.~\ref{sect:analysis_SP_i} (curved line) accurately predicts the observed values for $\tau=10$, as $\tau$ decreases the $\langle r_x \rangle$ values appear to only shift slightly to the right, preserving the underlying topology. Therefore, if $\tau$ is not too small (e.g., no S-P state was observed for $\tau=0.01$ for these parameters), then our analysis for large $\tau\gg\max\{1,\omega_0^{-1}\}$ can qualitatively predict the S-P state even when this assumption is violated.

%%%%%%%%%%%%%%%%%%%%%%%%%%%%%%%%%%%%%%%%%%%
\section{Discussion}\label{sect:conclusion}
%%%%%%%%%%%%%%%%%%%%%%%%%%%%%%%%%%%%%%%%%%%

%Due to the widespread need for theory describing complex systems with diversity \cite{multiplex,interactome,interconnected,power,motif,motif2}, 
We have introduced and analyzed a hybrid model consisting of interconnected Hopfield switches \cite{Hopfield} and Kuramoto phase oscillators \cite{Kuramoto1}, which respectively represent paradigmatic models for studying switch-like behavior
\cite{bool_gene,Switch_electronics,Chem_switch} and synchronization \cite{sync,Josephson,Lasers,FireFly,Animal,Clapping,Circadian}. 
In all-to-all networks with positive feedback, 
%Our development of this theoretical framework in Sec.~\ref{sect:analysis} was motivated and guided by recent research that proposed the hybrid model to study how network mutations in cell cycle regulation can lead to cancerous behavior \cite{elana}. 
%Their hybrid model recapitulated the system-wide dynamics of the yeast cell cycle, while demonstrating that small perturbations in the network topology result in cancer-like limitless activation of the cell cycle machinery.
%
%To simplify our initial development of analysis of a hybrid model, we have restricted our attention to all-to-all coupled switches and oscillators with positive feedback. Even so, 
rich dynamics were observed and analyzed, including three notable steady state solutions characterized by:
(i) incoherent oscillators and all switches permanently off (I-Off), 
(ii) synchronized oscillators and all switches permanently on (S-On), or 
(iii) synchronized oscillators and switches that periodically alternate between the on and off states (S-P). 
This latter case can be divided into cases where the average number of switches on remains fixed, but individual switches oscillate (when phase lags are uniformly distributed) and cases where the bulk of switches oscillate between on and off (when phase lags are identical).  Intermediate states are possible for different distributions of phase lags.  

In Sec.~\ref{sect:num} we numerically validated our results, highlighted their applicability outside of our assumptions, and explored transitions between these steady states (i-iii). 
Specifically, for sufficiently large systems, transitions between these states may be deterministically onset by the slow varying of a system parameter. These are well described by dynamic bifurcation theory \cite{DynBif} provided that the system is sufficiently large and that the parameter is varied sufficiently slow. 
%We also introduced state diagrams to summarize transitions and identify their reversibility. Importantly, it was suggested that topological changes occurring to such state diagrams may offer insight toward control mechanisms of complex systems. For example, we observed that a slight variation in one parameter caused the network topology of the state diagram to shift from one allowing hysteresis to one allowing only a cascade of irreversible transitions (e.g., as corresponds to a cancer-free cell cycle \cite{Novak}). 
%
For smaller system sizes we found that our system can spontaneously jump from the basin of attraction of one state to the basin of attraction of another due to finite-size fluctuations (which describe the discrepancy between asymptotic theory and systems with finite-size \cite{daido}). We note that similar spontaneous transitions were previously observed for systems of coupled oscillators \cite{vary_freq,macro}.

In summary, we have proposed and studied a hybrid system of coupled oscillators and switches, and have shown that it exhibits rich dynamics including multi-stability and hysteresis. This hybrid system was designed to serve as a simple example of a complex system with dynamical elements of different types, and thus several simplifying assumptions were made. 
In particular, two simplifications allowed us to neglect the effect of network topology in the present study: (i) the coupling between oscillators and switches was taken to be all-to-all, and (ii) we allowed the switches to affect the oscillators through an adaptive global coupling strength $k(t)$. 
%
%These particular simplifications were selected so that the topology of network-coupling, which is well known to significantly alter the dynamics of coupled switches or oscillators (e.g., by varying $K_0$ \cite{vary_net}), is not a factor determining the dynamics of our hybrid system, Eqs.~(\ref{eq:system1}-\ref{eq:system3}). 
%
If either or both of these assumptions are modified, then it is expected that more complicated dynamics will arise reflecting heterogeneities present in the network and/or switch dynamics (e.g., as observed in Ref.\cite{elana}). 
%
%functions between the switches and oscillators. While more general models tailored to specific applications could be studied, our model already demonstrates how the interplay between different dynamical units can result in novel phenomena.
%
Our model can therefore be used as a testbed to study the effect of heterogeneity and coupling network structure in collections of hybrid complex systems, and potentially to elucidate control mechanisms to alter their states. 
Finally, because we determined the stability of our system's dynamical states numerically, another fruitful direction of research includes the analysis of stability and classification of bifurcations. 

%%%%%%%%%%%%%%%%%%%%%%%%%%%%%%%%%%%%%%%%%%%%
\acknowledgements
%%%%%%%%%%%%%%%%%%%%%%%%%%%%%%%%%%%%%%%%%%%%

The work of D. T. and J. G. R. was supported by NSF Grant No. DMS-0908221. E. J. F was supported by NIH/NCI Grant CA141053. We also thank Matthew Francis for advice in the development of the hybrid model.

%%%%%%%%%%%%%%%%%%%%%%%%%%%%%%%%%%%%%%%%%%%%
~\\
\appendix

%\section{Linear stability of the IO state}\label{appendix:stable}
%%%%%%%%%%%%%%%%%%%%%%%%%%%%%%%%%%%%%%%%%%%%

%%%%%%%%%%%%%%%%%%%%%%%%%%%%%%%%%%%%%%%%%%%%
\section{The S-P state for identical phase lags}\label{appendix:SP_id}
%%%%%%%%%%%%%%%%%%%%%%%%%%%%%%%%%%%%%%%%%%%%

In this Appendix we provide analysis for the S-P state of our hybrid system, Eqs.~(\ref{eq:system1}-\ref{eq:system3}), for switches with identical phase lags $\beta_m=\beta$ for all $m$. As discussed in Sec.~\ref{sect:analysis_SP_i}, for large system size $N,M\to\infty$ and slow coupling adaptation $\tau\gg\max\{1,\omega_0^{-1}\}$, the S-P state corresponds to a system in which the oscillators synchronize while the switches turn on and off together in unison, causing $r_x(t)$ to periodically switch between 1 and 0. 
%Similar to the analysis provided in Sec.~\ref{sect:analysis_SP_u}, we will assume large system size $N,M\to\infty$ and slow coupling adaptation $\tau\gg\max\{1,\omega_0^{-1}\}$ and look to derive a consistency equation for this state.
%
Because this periodic oscillation occurs on a much faster timescale than the dynamics for the switches and coupling strength, analysis may be developed using a separation of time scales while considering the time-averaged variables $\langle r_x \rangle$, $\langle r_\theta \rangle$, and $\langle k \rangle$. 

We now develop a consistency equation for $\langle r_x \rangle$. Assuming that our system Eqs.~(\ref{eq:system1}-\ref{eq:system3}) is in the S-P state with $r_x(t)$ periodically switching between 0 and 1 at frequency $\omega_0$, one can show that $k(t)$ attains a trajectory of the form $k(t) = K\langle r_x \rangle+\mathcal{O}(T_0/\tau)$. It follows that for sufficiently large $\tau$, $k$ is approximately constant, $k=K\langle r_x\rangle$, and $r_\theta$ is given by Eq.~\eqref{eq:r_L} with $k=K\langle r_x\rangle$. With constant $k$ and $r_\theta$ and $\psi(t)=\omega_0 t +\psi(0)$, it remains to integrate Eq.~\eqref{eq:system1} for $x_m=x$ for all $m$ and fluctuating $r_x(t)$.

%use the assumption that a periodic steady state exists to directly integrate Eq.~\eqref{eq:identical_b}, and then use this result to develop the consistency equation given by Eq.~\eqref{eq:G}. 
We begin by separating $x(t)$ into two parts: a function $y(t)$ that is dependent on the average fraction of on switches, $\langle r_x\rangle$, and a function $z(t)$ that is piecewise-defined to account for fluctuations. Specifically, we let
\begin{equation} 
x(t) = y(t) + z(t),\label{eq:XXX}
\end{equation} 
where
\begin{eqnarray}
\frac{dy}{dt}  &=& -y-\eta + k^{x} \langle r_x \rangle+ k^{x\theta} \langle r_\theta \rangle\sin(\psi-\beta ) ,\label{eq:dy}\\
\frac{dz}{dt}   &=& -z+k^{x} \left\{\begin{array}{ccc} 1-\langle r_x \rangle &,&x(t)>0\\
-\langle r_x \rangle&,&x(t)\le 0 .
\end{array} \right. \label{eq:dz}
\end{eqnarray}
Note that adding the r.h.s. of the above equations recovers the r.h.s. of Eq.~\eqref{eq:system1}. As defined, because $x(t)$ is assumed to be periodic, $z(t)$ is necessarily periodic. 
Integration of Eq.~\eqref{eq:dy} yields steady state solutions given by
\begin{equation}
y(t) = U +V\sin(\omega_0 t -\beta-\delta), \label{y}
\end{equation}
where $U=\Big( k^{x}\langle r_x \rangle - \eta \Big)$, $V={k^{x\theta} \langle r_\theta \rangle}\cos(\delta) $, and $\delta=\arccos\left( {1}/{\sqrt{1+\omega_0^2}}\right)$.
Under the assumption that $z$ is periodic with the same period as $y$, $T_0=2\pi/\omega_0$, integration of Eq.~\eqref{eq:dz} leads to a piecewise-defined periodic solution 
\begin{widetext}
\begin{eqnarray}
z(t)  &=& \left\{\begin{array}{ccc}z_1e^{-\text{mod}\left(t-t_1,\frac{2\pi}{\omega_0}\right) } + k^{x}\Big(1-\langle r_x \rangle \Big)\left(1-e^{-\text{mod}\left(t-t_1,\frac{2\pi}{\omega_0}\right) }\right)  &,&\text{mod} \big(t-t_1,\frac{2\pi}{\omega_0} \big)\le t_2-t_1 \\
z_2e^{-\left[\text{mod}\left(t-t_1,\frac{2\pi}{\omega_0}\right)-(t_2-t_1)\right]} - k^{x} \langle r_x \rangle \left(1-e^{-\left[\text{mod}\left(t-t_1,\frac{2\pi}{\omega_0}\right)-(t_2-t_1)\right]}\right)   &,&\text{mod}\big(t-t_1,\frac{2\pi}{\omega_0} \big)\ge t_2-t_1 .\end{array} \right. \label{eq:Z}
\end{eqnarray}
\end{widetext}
Note that as defined, times $t\in\{t_1+ l \frac{2\pi}{\omega_0}\}$ for $l=0,1,\dots$ correspond to when $x(t)=0$, $dx(t)/dt>0$, and $z(t)$ attains its minimum value, $z_1$. On the other hand, times $t\in\{t_2+ l \frac{2\pi}{\omega_0}\}$ for $l=0,1,\dots$ correspond to when $x(t)=0$, $dx(t)/dt<0$, and $z(t)$ attains its maximum value, $z_2$. In other words, $x\ge0$ for $\text{mod}(t-t_1,2\pi/\omega_0)\in[0,t_2-t_1]$  while $x\le0$ for $\text{mod}(t-t_1,2\pi/\omega_0)\in[t_2-t_1,2\pi/\omega_0]$. Requiring that $z(t)$ is both periodic and continuous allows us to solve
\begin{eqnarray}
z_1 &=& -k^{x}\left(\langle r_x\rangle - \frac{e^{-2\pi/\omega_0(1-\langle r_x \rangle)} -e^{-2\pi/\omega_0} }
{1-e^{-2\pi/\omega_0} } \right) ,\label{eq:A6}\\
z_2 &=& -k^{x}\left( \langle r_x\rangle - \frac{1 -e^{-2\pi/\omega_0\langle r_x\rangle} }
{1-e^{-2\pi/\omega_0} } \right),\label{eq:A7}
\end{eqnarray}
where we have used that $\langle r_x \rangle=\frac{\omega_0}{2\pi}(t_2-t_1)$ by definition. 
In fact this definition may be used to write down a self-consistency equation for $\langle r_x \rangle$. Recalling that $t_1$ and $t_2$ were defined by the property $x(t_1)=x(t_2)=0$, we may use $y(t_1)=-z_1$ and $y(t_2)=-z_2$ to find
\begin{eqnarray}
t_1 &=& \omega_0^{-1}\left[{\delta-\arcsin\left( \frac{U+z_1}{V}\right)}\right] ,\label{eq:A8}\\
t_2 &=& \omega_0^{-1}\left[{\delta + \pi+\arcsin\left( \frac{U+z_2}{V}\right)}\right] .\label{eq:A9}
\end{eqnarray}
This leads to the consistency equation 
\begin{equation}
 G(\langle r_x\rangle) = \langle r_x\rangle - \frac{\omega_0}{2\pi}(t_2-t_1) =0,\label{eq:consistency2}
\end{equation}
where $t_1$ and $t_2$ depend implicitly on $\langle r_x \rangle$ through Eq.~\eqref{y} and Eqs.~(\ref{eq:A6}-\ref{eq:A9}).

%%%%%%%%%%%%%%%%%%%%%%%%%%%%%%%%%%%%%%%%%%%%%%%%
%%%%%%%%%%%%%%%%%%%%%%%%%%%%%%%%%%%%%%%%%%%%%%%%
%%%%%%%%%%%%%%%%%%%%%%%%%%%%%%%%%%%%%%%%%%%%%%%%
%%%%%%%%%%%%%%%%%%%%%%%%%%%%%%%%%%%%%%%%%%%%%%%%

\bibliographystyle{plain}

\end{document}